\documentclass[acmsmall, screen]{acmart}
\AtBeginDocument{%
  }

\usepackage{tcolorbox} % for summary box
\usepackage{fontawesome5} % for icons
\usepackage{algorithm}
\usepackage[noend]{algpseudocode}
\usepackage{listings}
\usepackage{xcolor}
\usepackage{colortbl}
\usepackage{verbatim}
\usepackage{booktabs}
\usepackage{array}
\usepackage{tabularx}
\usepackage{threeparttable}
\usepackage{multirow}
\usepackage{graphicx}
\usepackage{subcaption}
\usepackage{threeparttable} % 记得在导言区加入这个宏包
\AtBeginDocument{%
  \setlength{\abovedisplayskip}{4pt plus 2pt minus 2pt}%
  \setlength{\belowdisplayskip}{4pt plus 2pt minus 2pt}%
  \setlength{\abovedisplayshortskip}{3pt plus 1pt minus 1pt}%
  \setlength{\belowdisplayshortskip}{3pt plus 1pt minus 1pt}%
}

\newtcolorbox{boxK}{
    top=1pt,
    bottom=1pt,
    left=2pt,
    right=2pt,
    boxrule = 0pt,
    toprule = 0pt, % top rule weight
}

\begin{document}

\title{Porting Declarative UI to HarmonyOS: A Heuristic-guided Approach for lightweight LLMs}

\author{Kunwu Zheng}
\orcid{0009-0000-7047-1981}
\authornotemark[1]
\affiliation{%
  \institution{Shandong University}
  \city{Qingdao}
  \country{China}
}
\email{xiaozhengsdu2022@mail.sdu.edu.cn}

\author{Pengyu Xue}
\orcid{0009-0007-3395-9575}
\authornote{These authors contributed equally to this work.}
\affiliation{%
  \institution{Shandong University}
  \city{Qingdao}
  \country{China}
}
\email{xuepengyu@mail.sdu.edu.cn}

\author{Zhen Yang}
\orcid{0000-0003-0670-4538}
\authornote{Corresponding author.}
\affiliation{%
  \institution{Shandong University}
  \city{Qingdao}
  \country{China}
}
\email{zhenyang@sdu.edu.cn}

\author{Tianlong Chen}
\orcid{0009-0008-7750-5601}
\affiliation{%
  \institution{University of Jinan}
  \city{Jinan}
  \country{China}
}
\email{chentl@stu.ujn.edu.cn}

\author{Xiran Lyu}
\orcid{0009-0002-8861-8907}
\affiliation{%
  \institution{Shandong University}
  \city{Qingdao}
  \country{China}
}
\email{xiranlyu@mail.sdu.edu.cn}

\author{Peishi Lai}
\orcid{0009-0005-7728-9150}
\affiliation{%
  \institution{Shandong University}
  \city{Qingdao}
  \country{China}
}
\email{pacelai@mail.sdu.edu.cn}

\author{Mengying Zhao}
\orcid{0000-0002-1189-7787}
\affiliation{%
  \institution{Shandong University}
  \city{Qingdao}
  \country{China}
}
\email{zhaomengying@sdu.edu.cn}

\author{Junjie Chen}
\orcid{0000-0003-3056-9962}
\affiliation{%
  \institution{Tianjin University}
  \city{Tianjin}
  \country{China}
}
\email{junjiechen@tju.edu.cn}

\author{Yutian Tang}
\orcid{0000-0001-5677-4564}
\affiliation{%
  \institution{University of Glasgow}
  \city{Glasgow}
  \country{United Kingdom}
}
\email{Yutian.Tang@glasgow.ac.uk}

\author{Huizhi Zhang}
\orcid{0009-0004-8318-1624}
\affiliation{%
  \institution{Shandong University}
  \city{Qingdao}
  \country{China}
}
\email{chiparon@mail.sdu.edu.cn}

\author{Xianhang Li}
\orcid{0009-0006-8852-1215}
\affiliation{%
  \institution{Shandong University}
  \city{Qingdao}
  \country{China}
}
\email{xhli@mail.sdu.edu.cn}

\author{Linhao Wu}
\orcid{0009-0001-7624-156X}
\affiliation{%
  \institution{Peking University}
  \city{Beijing}
  \country{China}
}
\email{wulinhao26@stu.pku.edu.cn}

\author{Chengyi Wang}
\orcid{0009-0000-4153-6774}
\affiliation{%
  \institution{Shandong University}
  \city{Qingdao}
  \country{China}
}
\email{202300130150@mail.sdu.edu.cn}

\renewcommand{\shortauthors}{Kunwu Zheng et al.}

\begin{abstract}
   
   As an emerging operating system, HarmonyOS has a significant demand for software migration from platforms such as Android and iOS, where the User Interface (UI) translation accounts for a critical link. However, recent UI development has shifted to declarative paradigms, e.g., Kotlin Jetpack Compose (KJC) for Android, SwiftUI for iOS, and ArkUI for HarmonyOS, making prior translation approaches inapplicable because they target either backend logic or legacy imperative UIs. Thus, related research is limited.
   % As such, this paper targets ArkUI and proposes an automatic translation approach, namely \textit{ArkTrans}, to port UI files from Android and iOS to HarmonyOS.

  %\textit{ArkTrans} overcomes two salient challenges during the translation: (1) Programming Language (PL) unfamiliarity, and (2) the inability of generic information to substitute for the missing knowledge. Towards the first challenge, \textit{ArkTrans} heuristically constructs ArkUI skeletons by extracting metadata from source PL, thereby guiding LLMs' initial translation. %As for the second challenge, \textit{ArkTrans} executes empirically revealed post-fixing rules via pattern matching to repair most of the remaining syntactic errors. 
  %As for the second challenge, \textit{ArkTrans} performs compiler-guided iterative repair, leveraging compiler diagnostics to localize and fix the remaining syntactic errors. To examine the effectiveness of \textit{ArkTrans}, we construct a 100-sample parallel UI page translation benchmark from KJC/SwiftUI to ArkUI at the file level. %Extensive experiments demonstrate that LLMs with direct/one-shot prompting cannot translate a single compilable UI page. In contrast, 64.00-70.00\% \textit{ArkTrans}-translated files can be successfully compiled with high visual fidelity.   
  %Extensive experiments on the lightweight LLM show that LLM with direct prompting cannot translate a single compilable UI page with a 0\% compilation success rate, whereas \textit{ArkTrans} raises the compilation success rate to 70.00\% on KJC and 64.00\% on SwiftUI with high visual fidelity.
To systematically evaluate KJC/SwiftUI-to-ArkUI migration, we build a parallel benchmark of 100 KJC and 100 SwiftUI files across eight scenarios, plus 52 real-world UI pages from nine open-source apps. Evaluating nine direct-translation setups (including diverse LLMs and agents), we find Compilation Success Rate (CSR) spans only 0.00\%–35.00\%; common strategies (reference screenshots and compiler-guided repair) lift capable setups to 94.00\%–99.00\% CSR but fail on lightweight LLMs. By manually analyzing a 3,102-sample subset, deriving a regular-expression pattern for each failure family, and applying these patterns to generalize the resulting taxonomy to the full 29,593-sample set, we find residual errors rooted in (1) broken structure, (2) nonexistent APIs, and (3) missing resources (e.g., colors, strings, and icons of target projects).

Accordingly, we propose \textit{ArkTrans}, a declarative UI translation framework that maps these roots to deterministic structure supply, ArkUI knowledge and resource reference injection via a demonstration example, and iterative repair targeting residual errors, via four steps: Metadata Extraction, Skeleton Construction, LLM-driven Translation, and Compiler-Guided Iterative Repair. With Qwen3.6-27B, \textit{ArkTrans} raises CSR from 0.00\% to 70.00\% on KJC and 64.00\% on SwiftUI, achieving strong visual fidelity, and attains 65.63\%/80.00\% CSR on real-world pages, far outperforming the best common strategy at lower token cost.

\end{abstract}




\maketitle

\section{Introduction}
% 目前安卓和ios相关应用很多，相比之下，作为新兴os，harmonyos上软件支持不够丰富。因此针对harmonyos的软件自动化转译对于快速完善其软件生态很重要。
% 以往的研究大多关注于翻译软件的后端逻辑，如...，尽管取得了不少成就，但这些方法在移动端UI翻译上并不适用。当前主流移动端UI开发遵循一套声明式的编码范式和组件及其属性的嵌套式堆叠布局，而后端代码更关注复杂业务逻辑的处理。然而，面向harmonyos的UI翻译也存在一系列挑战。1. ArkUI相关代码量稀缺，LLM缺少训练语料，不了解arkui语法。（解决方案：中间表示辅助llm翻译） 2. 由于不了解arkui语法，进一步加剧了对源pl语法的复制现象。（解决方案:post-fixing） 
% 为了实现移动端UI的自动化翻译，本文提出ArkTrans。
As the mobile ecosystem continues to expand, Android and iOS dominate with a vast array of applications. In contrast, HarmonyOS, as an emerging operating system, suffers from a relative scarcity of software support. Automatically translating existing applications to HarmonyOS, therefore, presents a promising direction for rapidly enriching its software ecosystem. This paper focuses on mobile code translation from the perspective of the User Interface (UI), one of the most critical components of mobile applications \cite{qasim2018mobile}. However, the latest UI development has progressively shifted to the declarative programming paradigm in practice
% , where the interface is constructed through nested hierarchies of components and their attributes 
\cite{zhou2025declarui}. For example, Android applications use Kotlin Jetpack Compose (KJC) for UI development \cite{marchenko2023jetpack}, iOS applications adopt SwiftUI \cite{10522550}, while applications on HarmonyOS are developed with ArkUI \cite{zhou2025declarui}, which are three UI development frameworks of their respective PLs, i.e., Kotlin \cite{jemerov2017kotlin}, Swift \cite{mathias2016swift}, and ArkTS \cite{zhou2025porting}. Therefore, although prior research on automatic program translation has achieved notable success across functional to repository levels, their proposed approaches are ill-suited for translating mobile UIs, because they primarily focused on backend logic, such as converting server-side software with complex control flows and data manipulations \cite{yang2024exploring, 10.1145/3797099,ibrahimzada2025alphatrans,xue2025classeval}. Very few of the other research \cite{gao2024rule,gong2025uitrans} that lie in the mobile UI translation area remain within the scope of translating imperative UI, such as XML \cite{jackson2014introduction} and UIKit \cite{sadun2013ios}, leaving them powerless to apply to the latest declarative UI. Therefore, it is urgent to develop effective code translation tools for the latest mobile UI migration. 

To systematically evaluate the KJC/SwiftUI-to-ArkUI UI migration task, we first construct a benchmark at a cost of 200 person-hours, comprising a parallel set of 100 KJC and 100 SwiftUI files covering eight application scenarios such as e-commerce, smart home, and social media, along with 52 real-world UI pages collected from nine open-source mobile apps. Based on this benchmark, we evaluate nine direct-translation setups, including five LLM-only systems, three agentic coding harnesses, and CodeArts, the vendor's official AI-powered IDE plugin for HarmonyOS development. The results are all unsatisfactory: Compilation Success Rate (CSR) spans 0.00\%--35.00\%, and the gains from agent harnesses are inconsistent, e.g., Codex is slightly higher than its backbone on KJC but lower on SwiftUI. We then examine two common strategies, i.e., attaching a reference screenshot and performing compiler-guided iterative repair. Both lift capable models to near-perfect execution, with CSR reaching 94.00\%--99.00\% when combining both strategies, yet they leave the lightweight LLM below 26\%, because they only tell the LLM what the UI should look like and where the code goes wrong, but not how correct ArkTS should be written. Our failure analysis, which manually analyzes 3,102 samples, derives per-family regular-expression patterns, and applies them to generalize the resulting taxonomy to all 29,593 compiler diagnostics, further shows that the residual errors concentrate in three roots: broken structure, nonexistent APIs, and missing resources  (e.g., colors, strings, and icons of target projects).

% The above results reveal a key issue: the common strategies perform well on flagship models but are of little help for lightweight models with local deployment requirements, and generic information alone cannot compensate for their lack of ArkUI knowledge. 
Given the critical importance of lightweight LLMs for local deployment and privacy protection in both industry and academia, we propose \textit{ArkTrans} to enable lightweight LLMs to port declarative UIs to HarmonyOS.
% a declarative UI translation framework for KJC/SwiftUI-to-ArkUI migration. 
\textit{ArkTrans} maps the three failure roots to three design principles: structure should be supplied deterministically, ArkUI knowledge (e.g., API usage and resource references) should be injected through verified templates and a demonstration example, and the remaining errors should be left to iterative repair that starts from few errors. It implements these principles with four steps: Metadata Extraction (ME), Skeleton Construction (SC), LLM-driven Translation (LT), and Compiler-Guided Iterative Repair (CR). ME aims to extract metadata from source UI files and construct UI trees for maintaining necessary elements and layouts. The SC step synthesizes a deterministic skeleton, thereby finalizing the UI topology with partial ArkUI code while ensuring the maximum preservation of the metadata that cannot be migrated with heuristics. Subsequently, LT invokes an LLM to translate the above ArkUI skeleton into executable code with a specific example for guidance. Finally, CR compiles the generated code with the real hvigor compiler  (HarmonyOS's official build tool, which compiles ArkUI source)  in a minimal HarmonyOS project and feeds the diagnostics back to the LLM for iterative repair within a fixed iteration budget.

To examine the effectiveness of \textit{ArkTrans}, we evaluate it on the 200 benchmark files with the lightweight model Qwen3.6-27B as the backbone. \textit{ArkTrans} lifts the CSR from 0.00\% to 70.00\% on KJC and 64.00\% on SwiftUI. Regarding visual fidelity on KJC/SwiftUI, \textit{ArkTrans} achieves \textbf{Color Histogram Similarity} 45.16/41.40, \textbf{CLIP Similarity} 64.51/59.70, \textbf{Position Score} 49.60/49.07, \textbf{Size Score} 51.59/48.84, \textbf{Color Score} 43.17/40.47, and \textbf{Textual Integrity} 64.58/61.44, respectively. Ablation studies show that every component is necessary: removing the skeleton or iterative repair significantly reduces CSR, while removing the reference screenshot mainly affects color fidelity and, on SwiftUI, both CSR and color. On 52 real-world UI pages collected from nine open-source mobile apps, \textit{ArkTrans} attains 65.63\% CSR on KJC and 80.00\% on SwiftUI, whereas the best common strategy reaches only 12.50\% and 15.00\%  at 25,052/18,231 tokens per file, and ArkTrans does so at a lower token cost (17,423/6,520). These results demonstrate that ArkTrans turns the lightweight model into a usable UI translator and substantially outperforms the common strategies. The main contributions are fourfold:
%and \textit{ArkTrans} does so at a lower token cost (17,423/6,520). These results demonstrate that \textit{ArkTrans} turns the lightweight model into a usable UI translator and substantially outperforms the common strategies.
%(1) We construct the first KJC/SwiftUI-to-ArkUI translation benchmark, consisting of 100 file-level samples covering eight diverse application scenarios.
%(1) We construct the first KJC/SwiftUI-to-ArkUI translation benchmark, consisting of 100 parallel UI pages (200 files in total) covering eight diverse application scenarios.

(1) We construct the first KJC/SwiftUI-to-ArkUI benchmark of 100 parallel UI pages (200 files, eight scenarios), plus 52 real-world pages from nine open-source apps.

% (2) We conduct the first empirical study on KJC/SwiftUI-to-ArkUI translation, manually analyzing 3,102 samples and generalizing the resulting failure taxonomy to 29,593 samples, thereby summarizing a series of typical failures that guide the design of \textit{ArkTrans}.
(2) We conduct the first empirical study on KJC/SwiftUI-to-ArkUI translation, manually analyzing a 3,102-sample subset and generalizing the resulting failure taxonomy to the full 29,593-sample set, thereby summarizing typical failures that guide the design of \textit{ArkTrans}.
 
% We propose ArkTrans, a skeleton-guided and compiler-repair-enhanced code translation framework for porting declarative UIs to HarmonyOS, 
(3) Targeting lightweight LLMs, we propose ArkTrans, a skeleton-guided code translation framework with compiler-assisted repair, for porting declarative UIs to HarmonyOS. Accordingly, we establish a progressive evaluation system for both code executability and visual fidelity.

(4) We conduct extensive experiments and empirically prove the effectiveness of \textit{ArkTrans}. 
All code, benchmarks, and results are shown in \cite{ArkTrans:online}.

% 引用rq1-3的简短结论
\section{Motivation}
\label{Motivation}

\textbf{Direct migration fails across the board.} Our zero-shot evaluation, presented in Section~\ref{RQ1}, covers nine setups on 200 benchmark UI files. %including five LLMs called directly, three agentic coding harnesses, and the vendor's official IDE plugin. 
Compilation success caps at 35\%, agent harnesses rarely outperform the direct call to their own backbone, and the lightweight model Qwen3.6-27B produces nothing compilable at all.

\textbf{Common strategies are capability-gated.} 
% The two common strategies that require no task-specific engineering, 
Attaching a reference screenshot and performing compiler-guided iterative repair lift capable models to near-perfect execution, yet Section~\ref{RQ2} shows that they leave the lightweight model below 26\%. The reason is that both are generic information: they tell the model what the UI should look like and where the code goes wrong, but not how correct ArkTS should be written.

\textbf{Failures reduce to three roots.} Our failure analysis in Section~\ref{RQ3} classifies 29,593 compiler diagnostics across all strategies and finds the residual errors concentrate in three roots. (1) Broken structure is repaired in early rounds, yet whole-file rewrites keep re-introducing it. (2) Nonexistent APIs stem from an ArkUI knowledge gap that no diagnostic fills. (3) Missing resources stem from persistent hallucination that no strategy suppresses.

\textbf{Roots dictate three design principles.} Structure should be supplied deterministically rather than repaired; ArkUI knowledge, including which APIs exist and how standard resources are referenced, should be injected through verified templates and a demonstration example; % and resource references mapped to the target; 
and the mechanical residue should be left to iterative repair that starts from few errors. \textit{ArkTrans} implements these principles in four steps, described next, and lifts the same lightweight model from 0\% to 70\% compilation success on KJC and 64\% on SwiftUI, as demonstrated in Section~\ref{RQ4}.

% The replication package can be found in \cite{ArkTrans:online}.
\section{Approach}

In this section, we elaborate on the methodology and design motivations of \textit{ArkTrans}, a UI migration framework designed for KJC/SwiftUI-to-ArkUI translation. Its four stages implement the three design principles derived in Section~\ref{Motivation}.

\vspace{-0.5em}
\subsection{Overview}

\begin{figure}[htbp]
  \setlength{\abovecaptionskip}{0pt}
  \centering
  \includegraphics[width=0.95\textwidth]{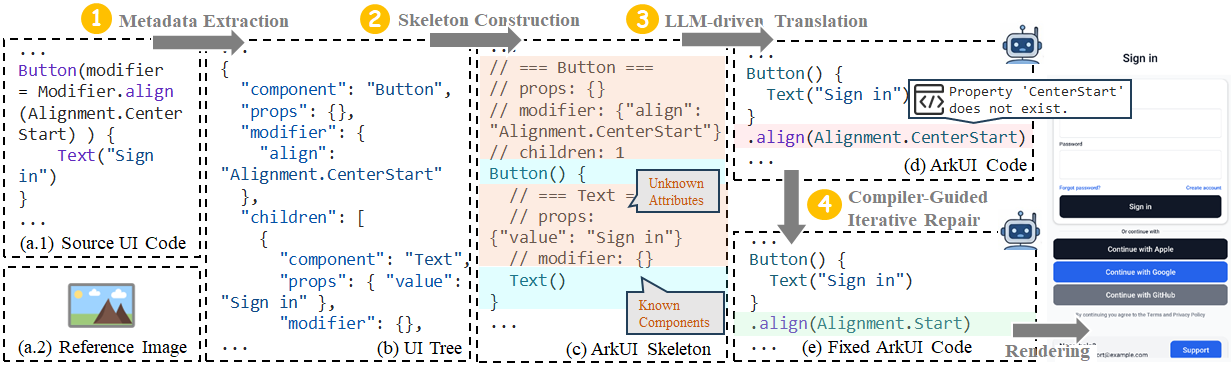}
  \caption{Workflow of \textit{ArkTrans}}
    \label{flow}
\end{figure}

Figure~\ref{flow} presents the workflow of \textit{ArkTrans} with a simplified translation example. \textit{ArkTrans} accepts UI files of source-PLs (KJC or SwiftUI) as inputs and outputs semantically equivalent ArkUI code. 
Overall, \textit{ArkTrans} consists of four main stages: 
(1) Metadata Extraction (ME): This stage parses the source UI file to extract its component hierarchies and associated properties for UI tree construction in JSON format. 
% The extracted data is then organized into a structured, hierarchical UI tree.
% (1)  Meta Data Extraction (ME), focuses on parsing the source UI files to extract their component hierarchy and properties. In this stage, a specialized parser scans the KJC or SwiftUI source code to identify the nesting relationsompiler-guided iterative Rephips of UI components. It also captures associated attributes, such as colors, alignment, and padding, along with code fragments that define local logic. 
% This extracted data is then organized into a structured JSON representation. By converting the raw source into a tree-based format, SD provides a structured data source that simplifies the subsequent mapping to ArkTS while preserving the original UI's structural and attribute-level details. 
% This extracted data is then organized into a tree-based format in JSON representation, thereby simplifying the subsequent mapping to ArkTS while preserving the original UI's structural and attribute-level details. 
(2) Skeleton Construction (SC): Using the above UI tree, \textit{ArkTrans} generates an ArkUI skeleton by performing cross-PL component mapping and establishes a layout constraint for follow-up semantic translation.
(3) LLM-driven Translation (LT): This stage invokes an LLM to populate the skeleton, guided by a one-shot skeleton-to-ArkUI example and the reference screenshot of the source UI, enabling the LLM to generate syntactically correct code while strictly adhering to the pre-defined layout topology. (4) Compiler-Guided Iterative Repair (CR): This stage compiles the output with the real hvigor compiler and feeds the diagnostics back to the LLM for repair within a fixed iteration budget.

% we implement a post-fixing engine to execute formal transformations on the output of Stage (3). The design of these rules is informed by our empirical analysis of common LLM hallucinations during UI migration. By systematically rectifying these residual artifacts, this stage resolves common syntax errors and ensures the final code strictly aligns with the HarmonyOS compiler requirements.

% Figure~\ref{result} presents a simplified example of the entire translation from (a) to (e), where a raw KJC code(a) is progressively transformed into a self-contained, compilable ArkUI code(e). The following subsections formally define each stage.
% A post-processing engine executes formal transformations on the output of step 3 to mitigate syntax hallucinations, where the model inadvertently retains source-lang1uage idioms.
% By rectifying residual artifacts and enforcing ArkTS-specific layout invariants, this stage ensures the final code aligns with the requirements of the HarmonyOS compiler to a great extent.

% \begin{figure*}[htbp]
%   \vspace{-1.5em}
%   \setlength{\abovecaptionskip}{0pt}
%   \centering
%   \includegraphics[width=0.9\textwidth]{flowchart.png}
%   \caption{Workflow of \textit{ArkTrans}.}
%     \label{flow}
% \end{figure*}

% 流程图占位
% \subsection{Step 1: Structural Distillation (SD)}
\subsection{Step 1: Metadata Extraction (ME)}

ME aims to hierarchically extract metadata from source UI files to construct a structural UI tree, which maintains the necessary elements (e.g., components and attributes) and UI layouts for follow-up semantic translation.
Each node in the tree is a tuple $\langle$ \texttt{comp}, \texttt{props}, \texttt{mod}, \texttt{children} $\rangle$. \texttt{comp}: the component type (e.g., \texttt{Button}, \texttt{Text}); \texttt{props}: a set of key-value pairs $(k_{p}, v_{p})$ representing named parameters passed to the component (e.g., $\{\text{value: "Confirm"}\}$); \texttt{mod}: a set of key-value pairs $(k_{m}, v_{m})$ representing chained modifier calls (e.g., $\{$align: Alignment.CenterStart$\}$); \texttt{children}: an ordered list of child nodes, preserving the nesting hierarchy of the original UI.

As shown in Algorithm~\ref {alg:distill}, we implement a recursive extraction process that leverages an Abstract Syntax Tree (AST) to identify UI patterns while filtering out non-UI logic.

\subsubsection{Entry Point Localization}

% Given the differing initialization mechanisms across platforms, the algorithm first identifies the logical starting point of the UI via static analysis. 
In declarative UI frameworks, initiating a UI page of mobile applications requires a segment of boilerplate configuration code as the entry for compilers to execute, which also serves as the root node of our entire UI tree. However, the syntax for these entry points varies significantly across platforms. To ensure the extraction process begins at the correct hierarchical level, we implement the \textsc{LocateEntryPoint} (Line 1).
For KJC, the algorithm parses the source code to an AST and locates top-level functions annotated with \texttt{@Composable} that possess no parameters, as these typically define the primary UI entry. For SwiftUI, the algorithm identifies \texttt{struct} declarations conforming to the \texttt{View} protocol and designates the internal \texttt{body} property as the starting point for recursive traversal. The \textsc{LocateEntryPoint} procedure isolates pure UI declarations from boilerplate configuration code, establishing a clear scanning boundary for subsequent extraction.

\subsubsection{Component Identification}

%Following entry point localization, the algorithm recursively traverses the AST to identify UI components. The identification logic (Lines 5–9) relies on a naming convention heuristic: an AST node $n$ is considered a potential UI component if it is a function call or initializer call that follows the PascalCase convention. This heuristic applies uniformly to both source frameworks:
Following entry point localization, the algorithm recursively traverses the AST to identify UI components using a naming convention heuristic (Lines 5–9): an AST node $n$ is considered a potential UI component if it is a function call or initializer call that follows PascalCase. This rule applies uniformly across both frameworks: in KJC, UI components are typically declared as @Composable functions and instantiated via PascalCase function calls (e.g., Button, Column); in SwiftUI, components are defined as structs conforming to the View protocol and instantiated via PascalCase initializer calls within the body (e.g., VStack, Text). 
% Although the underlying AST node types differ—function calls in Kotlin versus initializer calls in SwiftUI—both manifest syntactically as PascalCase identifiers, enabling a unified heuristic for component detection across the two frameworks.

To eliminate non-visual interference, we apply a component blacklist ($\mathcal{B}$) that prunes functional entities (Line 8). This blacklist includes PL-specific hooks such as \texttt{LaunchedEffect} in KJC, as well as SwiftUI's property wrappers, e.g.,  \texttt{@State}, \texttt{@Binding}, and lifecycle modifiers, e.g., \texttt{.onAppear}. The complete list of blacklisted tokens is provided in our repository for reference \cite{ArkTrans:online}. 
Besides, the algorithm maintains a set of custom component definitions $\Phi$, whose definitions are outside the entry point, via \textsc{CollectCusCompDefs} (Line 2). 
% Besides, the algorithm maintains a pre‑collected definition set $\Phi$ that contains all custom components whose definitions are outside the entry point (Line 2). 
If an identified component refers to a locally defined one in $\Phi$, the algorithm utilizes \textsc{FetchCompDef} (Line 11) to fetch its definition and parse its internal structure, ensuring that custom components are fully expanded into primitive UI elements.

\subsubsection{Attribute Extraction and Recursion}

Once a component node is confirmed, the algorithm harvests its associated properties and establishes ownership relationships (Lines 12--17). By iterating through the argument list, the algorithm captures chained calls and named parameters
into the \texttt{mod} and \texttt{props} fields, respectively.
% (e.g., \texttt{align} with the argument \texttt{CenterStart})
% (e.g., \texttt{"value": "Confirm"})
If an argument is not a literal or identifier, it may be a nested component. Then, a sub-recursive call will be triggered via \textsc{Extract} (Line 21), ensuring that the parent-child nesting relationships within the source code are faithfully reconstructed as branches in the hierarchical UI tree. Line 20's auxiliary function \textsc{Normalize} standardizes extracted values by removing quotes from strings, converting numeric strings to numbers, and resolving simple constant references.
% If an argument is itself a function call or a initializer call(Line 23), it is identified as a nested component, triggering a sub-recursive call to \textsc{Extract}. This strategy ensures that the parent-child nesting relationships within the source code are faithfully reconstructed as branches in the hierarchical UI tree.

% Concurrent with node identification, the algorithm performs granular extraction of property metadata. This stage aggregates styling and layout constraints from both explicit arguments and chained decoration methods into a unified pool. To ensure cross-platform compatibility, the \textsc{Normalize} operator resolves framework-specific values into standardized representations(e.g., from \textit{16.dp} to \textit{16})). If a property contains a nested component—such as a custom header passed as a parameter—the algorithm triggers a sub-recursive distillation call. This ensures that even complex, property-embedded UI structures are correctly encapsulated within the node's metadata before the final serialization.

\subsection{Step 2: Skeleton Construction (SC)}
\label{Step 2: Skeleton Construction (SC)}

Based on the aforementioned UI tree, the SC stage synthesizes a deterministic ArkUI code skeleton. Its primary objective is to finalize the UI topology while ensuring the maximum preservation of the metadata that cannot be migrated with heuristics.
% Based on the UI tree from Step 1, the SC stage synthesizes a deterministic ArkUI code skeleton. Its primary objective is to finalize the UI topology while ensuring the maximum preservation of the original component metadata. 
The detailed procedure is formalized in Algorithm \ref{alg:reify}.

% \begin{algorithm}[t]
% \caption{Skeleton Construction}
% \label{alg:reify}
% \footnotesize
% \begin{algorithmic}[1]
% \Require UI tree $\mathcal{T}_{\textbf{root}}$, 
% % $\mathcal{T}$ \{$comp$, $props$, $mod$, $children$\}
%  Component Mapping Dictionary $\mathcal{D}$
% \Ensure ArkUI code skeleton $S$
% \State $S \gets \text{BuildSkeleton}(\mathcal{T}_{\textbf{root}})$
% \Function{BuildSkeleton}{$n$}
%     \State $skeleton \gets \mathcal{D}[n.comp].template$
%     \For{$(k_{p}, v_{p}) \:\text{in}\: n.props$}
%         \If{$\mathcal{D}[n.comp]$ has property $k_{p}$}
%             \State $skeleton \gets \text{replace } k_{p} \text{ in } skeleton \text{ with } v_{p}$
%         \Else
%             \State $skeleton \gets skeleton + \text{`` // prop: $k_{p}$ = $v_{p}$''}$
%         \EndIf
%     \EndFor
%     \For{$(k_{m}, v_{m}) \:\text{in}\: n.mod$}
%         \If{$\mathcal{D}[n.comp]$ has modifier $k_{m}$}
%             \State $skeleton \gets skeleton + \text{``.$k_{m}$($v_{m}$)''}$
%         \Else
%             \State $skeleton \gets skeleton + \text{`` // modifier: $k_{m}$ = $v_{m}$''}$
%         \EndIf
%     \EndFor
%     \State $childSkeletons \gets []$
%     \For{$c \:\text{in}\: n.children$}
%         \State $childSkeletons.\text{append}(\text{BuildSkeleton}(c))$
%     \EndFor
%     \State $skeleton \gets skeleton.\text{replace}(\text{"\%s"}, childSkeletons)$
%     \State \Return $skeleton$
% \EndFunction

% \end{algorithmic}
% \end{algorithm}

% \subsubsection{Skeleton Synthesis}

First, it retrieves a component template for node $n$ from $\mathcal{D}$ (line 3). $\mathcal{D}$ is a mapping dictionary, defining a series of mappable components, properties, and modifiers from source PLs (KJC/SwiftUI) to their ArkUI counterparts; every entry is a manually verified ArkTS fragment, so the skeleton introduces only validated API and resource-reference forms. The specific mapping table can be found at \cite{ArkTrans:online}.
% summarized in Table~\ref{tab:mapping}, 
% which defines the mapping of components, properties, and modifiers from source PLs (KJC/SwiftUI) to their ArkUI counterparts.
Next, the algorithm iterates over the component's properties (lines 4–9): if a property is defined in $\mathcal{D}$, it is injected directly into the template; otherwise, it is preserved as a comment. Modifiers are processed similarly (lines 10–15): known modifiers become chained calls (e.g., \texttt{button().align().color()}), while unknown ones are kept as comments. After handling the current node, the algorithm recursively builds skeletons for all child nodes (lines 16–19). This yields a complete ArkUI skeleton with all known attributes translated and unknown ones annotated.

\begin{figure*}[t]
% \vspace{-1.0em}
\centering

% ===== 左栏：Algorithm 3 =====
\begin{minipage}[t]{0.48\linewidth}
\begin{algorithm}[H]
\caption{UI Tree Construction}
\label{alg:distill}
\scriptsize
\begin{algorithmic}[1]
\Require Source files $\mathcal{S}$, blacklist $\mathcal{B}$
\Ensure UI trees $\mathcal{T} = \{\text{comp}, \text{props}, \text{mod}, \text{children}\}$

\State $n_{\text{root}} \gets \text{LocateEntryPoint}(\mathcal{S})$
\State $\Phi \gets \text{CollectCusCompDefs}(\mathcal{S}, n_{\text{root}})$
\State $\mathcal{T} \gets \text{Extract}(n_{\text{root}}, \Phi, \mathcal{B})$

\Function{Extract}{$n$, $\Phi$, $\mathcal{B}$}
    \If{$n$ is not a function call or initializer call}
        \State \Return $\text{null}$
    \EndIf

    \State $name \gets \text{GetCallName}(n)$

    \If{$name \in \mathcal{B}$ \textbf{or} $name$ is not PascalCase}
        \State \Return $\text{null}$
    \EndIf

    \If{$name \in \Phi$}
        \State \Return $\text{Extract}(\text{FetchCompDef}(name,\Phi),\Phi,\mathcal{B})$
    \EndIf

    \State $\mathcal{T}.children \gets \mathcal{T}(\{name,\emptyset,\emptyset,[]\})$

    \For{$a \in \text{GetArguments}(n)$}
        \If{$\text{IsModifierCall}(a)$}
            \State $\mathcal{T}.\text{mod}[\text{ModName}(a)]
                \gets \text{ExtractValue}(a,\Phi)$
        \ElsIf{$\text{IsProperty}(a)$}
            \State $\mathcal{T}.\text{props}[\text{PropName}(a)]
                \gets \text{ExtractValue}(a,\Phi)$
        \EndIf
    \EndFor

    \State \Return $\mathcal{T}$
\EndFunction

% \Function{ExtractValue}{$a,\Phi$}
%     \State \Return $\text{Normalize}(a,\Phi)$ \textbf{if} $a$ is literal or identifier;
%     $\text{Extract}(a,\Phi,\mathcal{B})$ otherwise
% \EndFunction

\Function{ExtractValue}{$a,\Phi$}
    \State \Return $\text{Normalize}(a,\Phi)$ \quad \textbf{if} $a$ is literal or identifier
    \State \Return $\text{Extract}(a,\Phi,\mathcal{B})$ \quad \textbf{otherwise}
\EndFunction

\end{algorithmic}
\end{algorithm}
\end{minipage}%
\vspace{-0.5em}
\hfill
% ===== 右栏：Algorithm 4 =====
\begin{minipage}[t]{0.48\linewidth}
\begin{algorithm}[H]
\caption{Skeleton Construction}
\label{alg:reify}
\scriptsize
\begin{algorithmic}[1]
\Require UI tree $\mathcal{T}_{\textbf{root}}$, Component Mapping Dictionary $\mathcal{D}$
\Ensure ArkUI code skeleton $S$

\State $S \gets \text{BuildSkeleton}(\mathcal{T}_{\textbf{root}})$

\Function{BuildSkeleton}{$n$}
    \State $skeleton \gets \mathcal{D}[n.comp].template$

    \For{$(k_p,v_p) \in n.props$}
        \If{$\mathcal{D}[n.comp]$ has property $k_p$}
            \State $skeleton \gets \text{replace } k_p
                \text{ in } skeleton \text{ with } v_p$
        \Else
            \State $comment \gets \text{`` // prop: } k_p \text{ = } v_p \text{''}$
            \State $skeleton \gets skeleton + comment$
        \EndIf
    \EndFor

    \For{$(k_m,v_m) \in n.mod$}
        \If{$\mathcal{D}[n.comp]$ has modifier $k_m$}
            \State $skeleton \gets skeleton + \text{``.$k_m$($v_m$)''}$
        \Else
            \State $comment \gets
                \text{`` // modifier: } k_m \text{ = } v_m \text{''}$
            \State $skeleton \gets skeleton + comment$
        \EndIf
    \EndFor

    \State $childSkeletons \gets []$

    \For{$c \in n.children$}
        \State $childSkeletons.\text{append}(\text{BuildSkeleton}(c))$
    \EndFor

    \State $skeleton \gets
        skeleton.\text{replace}(\text{"\%s"}, childSkeletons)$
    \State \Return $skeleton$
\EndFunction

\State \Return $S$

\end{algorithmic}
\end{algorithm}
\end{minipage}%
% \vspace{-0.5em}
\end{figure*}

\subsection{Step 3: LLM-driven Translation (LT)}

In the LT stage, \textit{ArkTrans} invokes an LLM to translate the ArkUI skeleton into executable code. To achieve this, we construct a structured prompting strategy and an associated skeleton-to-ArkUI example for specific guidance. 
% This process relies on a structured prompting strategy to ensure that the LLM functions as a deterministic translator rather than a generative agent.

% \subsubsection{Prompt Structure}
% Figure~\ref{fig:prompt_design} lists the specialized prompt in \textit{ArkTrans}. 
The prompt for skeleton-to-ArkUI translation is divided into three functional segments: (1) the system role; (2) the contextual data, containing the aforementioned ArkUI skeleton and the reference screenshot of the source UI, which anchors the intended layout, colors, spacing, and text content; and (3) the one-shot example described below. The specific prompt can be found in \cite{ArkTrans:online}.
% This prevents the model from invoking non-existent APIs or mixing the source syntax with the target syntax. 
% By requiring the LLM to replace the comments with specific implementations, the prompt ensures that every visual property is anchored to its original location in the UI hierarchy, effectively preventing the hallucination of non-existent APIs.

% \subsubsection{One-shot Learning Example}

Although the skeleton above has been partially transformed into ArkUI code via heuristics as mentioned in Section \ref{Step 2: Skeleton Construction (SC)}, it still has lots of annotated elements of the source PL that need to be translated.   
To familiarize LLMs with programming ArkUI code according to the associated skeleton, we craft a demonstration example (available in our repository \cite{ArkTrans:online}) to guide LLMs for in-context learning.
% that serves as an in-context learning reference for LLMs.
% The need for such an example stems from fundamental syntactic differences between the source frameworks and ArkUI, as illustrated in Figure~\ref{fig:framework_diff}. 
The example covers most kinds of syntactic discrepancies between the source PLs and ArkUI, including property assignment, modifier chaining, layout property expansion, component naming, state management, and resource referencing; notably, many resources whose names LLMs invent are in fact available from the standard library with a correct reference syntax, which the example demonstrates in context, whose failure families are analyzed in Section~\ref{RQ3}.

\subsection{Step 4: Compiler-Guided Iterative Repair (CR)}

CR settles the mechanical residue that prompts alone cannot guarantee to eliminate. It first compiles the output of LT with the real hvigor compiler in a minimal HarmonyOS project. If compilation fails, CR feeds the diagnostics back to the LLM, which rewrites the complete file; the loop repeats until the code compiles or the iteration budget of $N$ rounds is exhausted. What the LLM receives is the raw message of the failed build, i.e., the error code, message, and file location of every reported problem, together with the current code. % no richer compiler capability, such as API documentation lookup, code completion, or automated quick fixes, is exposed to the model. 
We set $N = 5$ because Section~\ref{RQ3} shows that both the error count and the compilation success rate plateau well in the fifth round.

\section{Evaluation Design}

% This section presents the associated design for experimentation.
% , such as research questions, datasets, baselines, metrics, LLM selection, and implementation details.

\subsection{Research Questions}

%To comprehensively evaluate the efficacy of our framework, we formulate four distinct research questions.

%\textbf{RQ1 (Direct Migration Baseline)} focuses on assessing the inherent capabilities and limitations of SOTA LLMs in direct UI code translation from KJC and SwiftUIUI to ArkUI, identifying characteristic failure modes in unguided scenarios.

%\textbf{RQ2 (Comparative Effectiveness)} evaluates the performance gains achieved by our complete pipeline compared to the direct prompting baselines identified in RQ1.

%\textbf{RQ3 (Ablation Study))} investigates the individual contributions of our core components by incrementally introducing the ArkUI code skeleton in SR, the one-shot learning example in ST, and formal alignment in FA.

%\textbf{RQ4 (Cross-Model Performance)} verifies the framework's robustness across diverse large language models, ensuring that our methodology remains effective regardless of the underlying backbone architecture.

We propose Research Questions (RQs) below to comprehensively evaluate \textit{ArkTrans} in the context of cross-platform UI migration.

\textbf{RQ1 (Capability of Direct Translation): How far can existing LLMs and agent
frameworks translate KJC/SwiftUI code to ArkUI without dedicated guidance?} We evaluate nine setups in three categories: five LLMs called directly spanning flagship to lightweight models, three agentic coding harnesses, and the vendor's official IDE plugin. All setups receive an identical zero-shot prompt and translate each of the 200 benchmark files in a single shot. 
% We measure compilation success rate and visual fidelity, with failed compilations scored zero on all visual metrics.

% , categorizing errors into structural retention and syntactic residue. It then evaluates the improved translation using the skeleton and one-shot example, quantifying the remaining errors that motivate the design of our rule-based post-fixing stages.

% \textbf{RQ1 (Overall Performance): How effective is \textit{ArkTrans} in migrating UI pages to ArkUI compared to Direct Prompting prompting?}
\textbf{RQ2 (Performance of Common Strategies): How much do reference images and compiler iterative repair help LLMs and agents migrate UI code to ArkUI?} We augment the zero-shot setting of RQ1 with two common strategies that require no task-specific engineering: (1) \textit{Reference Image}, attaching the rendered screenshot of the source UI to the prompt, and (2) \textit{Compiler Iterative Repair}, feeding real hvigor compiler diagnostics back for repair, under an iteration budget of up to five repair rounds.

\textbf{RQ3 (Failure Mechanism): Why do the common strategies fail on the lightweight model?} We replay all four configurations of RQ2 on Qwen3.6-27B, a lightweight open-weight model, from their complete traces, and classify every compiler diagnostic into a two-level taxonomy of five families. The iteration runs cover five repair rounds with full hvigor diagnostics.

% \textbf{RQ4 (Our Method): How far does skeleton-guided migration lift the lightweight model?} 
\textbf{RQ4 (Our Method): How far does ArkTrans lift the migration capability of lightweight LLMs?} 
We instantiate the remedies dictated by RQ3 as \textit{ArkTrans} and evaluate on the same lightweight model. Besides, we report its overall performance on the 200-file benchmark, component ablation, and a transfer evaluation on 52 UI pages from nine real-world open-source mobile apps.
%: metadata extraction, deterministic skeleton construction, a single LLM translation call conditioned on the skeleton together with the one-shot ArkTS example and the reference screenshot, and a compiler-guided iterative repair (CR) loop over real hvigor diagnostics within a fixed budget. 

\subsection{Benchmark}
To our knowledge, there is no existing benchmark for assessing KJC/SwiftUI-to-ArkUI translation. In addition, files are the primary compilation unit in UI pages. Thus, we manually construct a file-level UI migration benchmark in this work, which costs 200 person-hours. In the field of code translation, the construction of high-quality parallel benchmarks typically follows the paradigm of starting from a calibration PL and manually extending it to target PLs \cite{yang2024exploring, xue2025classeval,xue2026translibeval}. To construct the benchmark, three software engineers, each with over three years of professional mobile development experience, manually implemented various UI effects derived from open-source Android projects using KJC. These implementations were then manually ported to SwiftUI to create functionally equivalent code files. Each implementation was engineered as a standalone page file with zero dependencies on external files or third-party libraries.

\begin{table}[htbp]

\centering
\setlength{\abovecaptionskip}{0.1cm}
\begin{threeparttable}
\caption{Statistical summary of the benchmark and real-world pages}
\label{tab:code_comparison}
\footnotesize
\renewcommand{\arraystretch}{0.92}%
\setlength{\tabcolsep}{0pt}
\begin{tabular*}{\linewidth}{@{\extracolsep{\fill}} llcccccccc @{}}
\toprule
\multirow{2}{*}{\textbf{Category}} & \multirow{2}{*}{\textbf{\#}} & \multicolumn{4}{c}{\textbf{KJC}} & \multicolumn{4}{c}{\textbf{SwiftUI}} \\ \cmidrule(lr){3-6} \cmidrule(l){7-10}
 & & \textbf{TK} & \textbf{CP} & \textbf{LOC} & \textbf{DEP} & \textbf{TK} & \textbf{CP} & \textbf{LOC} & \textbf{DEP} \\ \midrule
E-commerce   & 15 & 2362.1 & 31.9 & 280.7 & 8.4 & 1900.8 & 25.1 & 205.6 & 8.1 \\
Finance      & 13 & 2653.2 & 30.1 & 272.2 & 7.9 & 2280.4 & 27.4 & 241.5 & 8.5 \\
Lifestyle    & 16 & 2443.4 & 31.0 & 304.4 & 7.8 & 1886.8 & 31.7 & 210.6 & 7.8 \\
Smart Home   & 11 & 2486.6 & 57.6 & 334.8 & 8.9 & 2219.9 & 48.0 & 250.7 & 8.3 \\
Productivity & 14 & 2326.6 & 32.5 & 265.4 & 8.1 & 2191.5 & 33.4 & 233.2 & 8.9 \\
Education    & 7  & 2724.3 & 28.1 & 294.0 & 7.7 & 2080.3 & 40.4 & 234.7 & 10.1 \\
Social Media & 14 & 2284.9 & 25.6 & 254.4 & 7.6 & 1690.1 & 25.8 & 188.7 & 8.8 \\
Utilities    & 10 & 2746.5 & 48.2 & 346.9 & 8.6 & 2043.4 & 40.6 & 227.8 & 8.6 \\
\textbf{Overall} & \textbf{100} & \textbf{2474.7} & \textbf{34.9} & \textbf{291.1} & \textbf{8.1} & \textbf{2021.0} & \textbf{32.9} & \textbf{221.8} & \textbf{8.5} \\
\midrule
\end{tabular*}
% \vspace{1pt}
\begin{tabular*}{\linewidth}{@{\extracolsep{\fill}} lcccccclcccccc @{}}
\multicolumn{14}{l}{\textit{Pages from real-world mobile apps}} \\
\cmidrule(lr){1-7} \cmidrule(l){8-14}
\textbf{Project} & \textbf{Sou. PL} & \textbf{\#} & \textbf{TK} & \textbf{CP} & \textbf{LOC} & \textbf{DEP} & \textbf{Project} & \textbf{Sou. PL} & \textbf{\#} & \textbf{TK} & \textbf{CP} & \textbf{LOC} & \textbf{DEP} \\ \midrule
notes-app    & KJC & 8  & 7396.8 & 94.2  & 882.1 & 10.0 & clean-pad & SwiftUI & 14 & 959.4  & 10.7 & 136.7 & 4.9 \\
AudioNote    & KJC & 6  & 2541.3 & 29.3  & 314.2 & 6.7  & MovieApp  & SwiftUI & 3  & 1904.3 & 7.7  & 200.7 & 4.0 \\
EasyNotes    & KJC & 4  & 1595.2 & 23.8  & 200.8 & 6.8  & Weather   & SwiftUI & 3  & 870.0  & 9.7  & 128.3 & 6.0 \\
Grit         & KJC & 10 & 7140.6 & 34.0  & 936.7 & 6.0  &           &        &    &        &      &       &     \\
MpesaUiClone & KJC & 1  & 4564.0 & 100.0 & 573.0 & 14.0 &           &        &    &        &      &       &     \\
ToDo\_App    & KJC & 3  & 4443.7 & 50.0  & 633.3 & 10.0 &           &        &    &        &      &       &     \\ \midrule
\textbf{Overall} & \textbf{KJC} & \textbf{32} & \textbf{5315.8} & \textbf{50.5} & \textbf{674.5} & \textbf{7.8} & \textbf{Overall} & \textbf{SwiftUI} & \textbf{20} & \textbf{1087.8} & \textbf{10.1} & \textbf{145.1} & \textbf{5.0} \\
\bottomrule
\end{tabular*}
\begin{tablenotes}
    \footnotesize
    \item \textit{Note:} \textbf{\#} denotes the number of tasks; \textbf{TK}, \textbf{CP}, \textbf{LOC}, and \textbf{DEP} represent the \textbf{average} values of Tokens, Number of UI Components, Lines of Code, and Maximum Nesting Depth, respectively.
\end{tablenotes}
\end{threeparttable}

\end{table}

To ensure that the translated SwiftUI code achieves faithful consistency with the original Kotlin calibration in terms of both functionality and visual presentation, we introduce a rigorous verification mechanism. Beyond manual code review, each translation pair is executed, and its rendered screenshot is compared against the original KJC implementation; the pair is formally included in the benchmark only if all six visual similarity metrics defined in Section \ref{Evaluation Metrics} (CLIP, CH, Pos, Size, Color, and Text) exceed 90.00, and failing samples are iteratively calibrated and re-translated. Table \ref{tab:code_comparison} provides a comprehensive breakdown of the benchmark, including per-domain distribution and structural metrics for the 100 high-quality parallel UI page files. Notably, E-commerce (15\%), Lifestyle (16\%), and Productivity (14\%) constitute nearly half of the samples, ensuring the benchmark covers the most representative interaction patterns in current mobile applications. For evaluation, all 200 files are used throughout RQ1--RQ4. 

% \begin{table*}[htbp]
% \vspace{-1em}
% \centering
% \setlength{\abovecaptionskip}{0.1cm}
% \caption{Composition of the 52 real-world UI pages}
% \label{tab:realworld_pages}
% \footnotesize
% \begin{tabular*}{\textwidth}{@{\extracolsep{\fill}}lcccccc|ccc@{}}
% \toprule
%  & \multicolumn{6}{c|}{\textbf{KJC (6 apps)}} & \multicolumn{3}{c}{\textbf{SwiftUI (3 apps)}} \\
%  & notes-app & AudioNote & EasyNotes & Grit & Mpesa & ToDo & clean-pad & MovieApp & Weather \\
% \midrule
% Pages & 8 & 6 & 4 & 10 & 1 & 3 & 14 & 3 & 3 \\
% TK & 7,396.8 & 2,541.3 & 1,595.2 & 7,140.6 & 4,564.0 & 4,443.7 & 959.4 & 1,904.3 & 870.0 \\
% LOC & 882.1 & 314.2 & 200.8 & 936.7 & 573.0 & 633.3 & 136.7 & 200.7 & 128.3 \\
% \bottomrule
% \end{tabular*}
% \par\vspace{2pt}
% \footnotesize
% \flushleft
% \vspace{-2.0em}
% \end{table*}

To probe performance beyond this curated benchmark, RQ4 additionally collected 52 UI pages from nine real-world open-source mobile apps hosted on GitHub, six Android apps developed in KJC and three iOS apps in SwiftUI. We favored repositories whose UI pages carry light backend coupling, so that individual pages could be extracted with minimal adaptation, and detached each selected page from project-specific dependencies into a self-contained file that compiles and renders in isolation, following the standalone-page setting of the benchmark (statistics in Table~\ref{tab:code_comparison}). The complete list and the provenance of each page are provided in our repository \cite{ArkTrans:online}.

\subsection{Evaluation Metrics}
\label{Evaluation Metrics}
Previous research \cite{gao2024rule, gong2025uitrans} on UI migration has largely remained at the level of measuring compilation success rates or overall visual similarity, lacking a systematic evaluation for fine-grained elements against the complicated UI pages. Inspired by the UI code generation area \cite{si2025design2code}, we adapt a progressive evaluation system spanning two aspects: code executability and visual fidelity. In particular, the latter is further partitioned into global and local metrics for fine-grained analysis. For all the aforementioned metrics, higher values indicate superior migration performance, and all visual-fidelity scores are normalized to a 0--100 scale, with files that fail to compile scored zero on all visual metrics.
% The evaluation focuses on two core dimensions: code executability and visual fidelity, 
% the practical executability of the generated code and its visual alignment with the source UI. We employ a suite of seven metrics to provide a granular diagnostic assessment of the migrated components.

\subsubsection{Code Executability}

\textbf{Compilation Success Rate (CSR)}: 
%This metric computes the ratio of translated samples that can be successfully compiled within the target HarmonyOS environment. 
This metric computes the ratio of translated samples that can be successfully compiled within the HarmonyOS environment at the file level.
% It measures the capability to adhere to the strict syntactic rules and API constraints of ArkUI, ensuring the output is a deployable application rather than a fractured code snippet. 
CSR is defined as follows:

$$
\mathrm{CSR} = \frac{\sum_{k=1}^{N} cs(\hat{y}_k)}{N}, \text{ where } cs(\hat{y}_k) =
\begin{cases}
1 & \text{if Compile}(\hat{y}_k) \to \text{success} \\
0 & \text{if Compile}(\hat{y}_k) \to \text{error}
\end{cases}
$$

where $N$ denotes the total number of samples (files) and $\hat{y}_k$ represents the $k$-th translated ArkUI file.

\subsubsection{Visual Fidelity}

To holistically assess visual similarity, we employ both global and local metrics.

\paragraph{(1) Global Metrics}

\textbf{CLIP Similarity (CLIP)}: We measure high-level semantic alignment using a pre-trained vision-language model, namely CLIP \cite{radford2021learning}. For clarity, we use the model name as an abbreviation for this metric.

$$
\mathrm{CLIP} = \frac{E(I_R) \cdot E(I_G)}{\|E(I_R)\| \|E(I_G)\|}
$$

where $I_R$ and $I_G$ denote the reference and generated UI pages, respectively. $E(\cdot)$ denotes using CLIP-ViT-B/32 to encode UI pages as embeddings for similarity comparison.

\textbf{Color Histogram Similarity (CH)}: We quantify global tonal consistency by the intersection of normalized 48-bin HSV \cite{sural2002segmentation} histograms:

$$
\mathrm{CH} = \sum_{k=1}^{48} \min(H_{R}(k), H_{G}(k))
$$

where $H_R$ and $H_G$ are the HSV histograms of $I_R$ and $I_G$.

\paragraph{(2) Local Metrics}

Considering the complicated UI designs, we propose a multi-strategy block extraction pipeline using diverse image processing techniques, thereby improving the extraction accuracy for follow-up block-wise evaluation. Specifically, (1) we employ MSER \cite{matas2004robust} to detect small containers. (2) To capture large-scale containers and layout boundaries, we utilize Canny edge detection \cite{canny1986computational} followed by morphological closing \cite{haralick1987image}. This combination bridges fragmented edge pixels into closed structural units, preventing the omission of subtle layout frames. (3) Color quantization \cite{heckbert1982color} is concurrently applied to segment the blocks with substantial chromatic differences. All candidate blocks undergo Non-Maximum Suppression (NMS) \cite{neubeck2006efficient} with an empirically determined IoU threshold of 0.6 to prune redundant detections. The above process yields the final block sets $R = \{r_1, ..., r_m\}$ from the reference UI ($I_R$) and $G = \{g_1, ..., g_n\}$ from the generated UI ($I_G$), where $m$ and $n$ are the numbers of detected blocks in each image.
We establish optimal matches $M = \{(r, g)\}$ using the Jonker-Volgenant algorithm \cite{malkoff1997evaluation} to minimize a joint cost of spatial and chromatic distances. 

To ensure the evaluation aligns with human visual perception, we propose an area-weighted scoring mechanism that prioritizes visually larger blocks. First, we follow \cite{si2025design2code} to introduce the Bilateral Area Coverage ($C$) for global quality constraint, which simultaneously penalizes component omissions and hallucinations during LLM generation by normalizing the matched area of block sets from both reference and generated UI components against their total block area:

% to provide a global quality constraint. Unlike standard detection metrics that primarily focus on object recall, $C$ is designed to simultaneously penalize component omissions and hallucinations during LLM generation
% by normalizing the matched area against the total area of both reference and generated sets:
\vspace{-0.8em}
$$
C = \frac{\sum_{(r,g) \in M} (A(r) + A(g))}{\sum_{r \in R} A(r) + \sum_{g \in G} A(g)}
$$

where $A(r)$ and $A(g)$ denote the areas (in pixels) of blocks $r$ and $g$, respectively. Subsequently, we propose to employ the reference area $A(r)$ as a weighting factor for each pair of blocks in the follow-up score calculation, thereby ensuring that larger blocks account for more proportions to final scores. Each local metric is defined as follows:

% We employ the average area $\bar{A}_{r,g} = \frac{A(r) + A(g)}{2}$ as a local weighting factor for all fine-grained metrics. 
% This ensures that deviations in larger components impact the final score more than smaller ones, thereby reflecting the visual hierarchy inherent in UI design. The specific metrics are defined as follows:

\textbf{Position Score (Pos)}: this metric measures the area-weighted spatial deviation distance among all $(r, g)$ in $M$. The lower the deviation distance, the higher the \textbf{Pos}, meaning the higher-quality of translated code. The score is defined as:
% Spatial precision using Gaussian-decayed center distance between matched blocks:

$$
\mathrm{Pos} = C \times \frac{\sum_{(r,g) \in M} e^{-2 \cdot d_{pos}(r,g)} \cdot A(r)}{\sum_{r \in M} A(r)}
$$

where $d_{pos}(r,g) = \sqrt{(c_x^r-c_x^g)^2 + (c_y^r-c_y^g)^2}$ is the Euclidean distance between the normalized centers $(c_x, c_y) \in [0,1]^2$ of blocks $r$ and $g$. The exponential term decreases as the center distance increases, assigning higher scores to better-aligned blocks.
% The deviation score follows a Gaussian decay paradigm \cite{izrailev2006return}.  

\textbf{Size Score (Size)}: This metric evaluates area-weighted size fidelity through normalized width and height offsets. The score is defined as:

$$
\mathrm{Size} = C \times \frac{\sum_{(r,g) \in M} (1 - |w_r - w_g| - |h_r - h_g|) \cdot A(r)}{\sum_{r \in M} A(r)}
$$

where $w_r, h_r$ and $w_g, h_g$ represent the width and height proportions of blocks $r$ and $g$ relative to the whole page size ($W, H$). This proportional normalization ensures a resolution-invariant evaluation, preventing scores from being biased by differing image scales.

\textbf{Color Score (Color)}: This metric evaluates area-weighted chromatic accuracy via HSV histogram intersection for all matched pairs $(r, g) \in M$. The score is defined as:

$$
\mathrm{Color} = C \times \frac{\sum_{(r,g) \in M} \left( \sum_{k=1}^{K} \min(h_r(k), h_g(k)) \right) \cdot A(r)}{\sum_{r \in M} A(r)}
$$

where $h_r$ and $h_g$ represent the color distribution ratios (normalized HSV histograms) of blocks $r$ and $g$, and $K$ denotes the total number of bins in the color space. Each bin $k \in \{1, \dots, K\}$ acts as a specific color category; the term $\min(h_r(k), h_g(k))$ calculates the common overlap of these color categories between the two blocks.

\textbf{Textual Integrity (Text)}: 
We identify textual blocks and extract their content using EasyOCR \cite{easyocr2020} from matched blocks $(r, g) \in M$. On the basis of only textual block pairs $M_{text} \subseteq M$, we define Text below:
% Content accuracy within text regions using EasyOCR \cite{easyocr2020}. This metric is computed only on matched text block pairs $M_{text} \subseteq M$:

$$
\mathrm{Text} = C_{text} \times \frac{\sum_{(r,g) \in M_{text}} s_{text}(r,g) \cdot A(r)}{\sum_{r \in M_{text}} A(r)}
$$

where $s_{text}(r,g) = \frac{2 \cdot |\text{OCR}(r) \cap \text{OCR}(g)|}{|\text{OCR}(r)| + |\text{OCR}(g)|}$ is the character-level Dice coefficient between the extracted text strings, and $C_{text}$ is the bilateral area coverage computed exclusively on textual blocks.
\begin{comment}

\subsection{LLM Selection}
For RQ1 and RQ2, we employ GPT-5.2 as our primary backbone model due to its state-of-the-art reasoning capabilities. To further investigate the generalizability of \textit{ArkTrans} in RQ3, we extend our evaluation to four representative LLMs with diverse architectural backgrounds:
(1) \textbf{GPT-5.2}, a flagship model recognized for its advanced logical inference and cross-paradigm code understanding \cite{openai};
(2) \textbf{GLM-5}, a bilingual optimized model that demonstrates robust performance in structured data mapping and component alignment \cite{zhipuai2024glm5};
(3) \textbf{DeepSeek-V3.2}, an open-source model widely noted for its specialized aptitude in structural inference and low-resource code generation tasks \cite{deepseek}; and
(4) \textbf{Kimi-K2-Turbo}, which excels in long-context comprehension and following nuanced instructions for complex UI layouts \cite{moonshot2024}.
\end{comment}

\subsection{Implementation Details}
\label{Implementation Details}

% The nine setups of RQ1 consist of five LLMs called directly (\textit{i.e.}, GPT-5.5 \cite{openai}, Claude-Sonnet-4.6 \cite{anthropic}, DeepSeek-V4-Pro \cite{deepseek}, GLM-5.2 \cite{zhipuai2024glm5}, and the lightweight open-source Qwen3.6-27B \cite{qwen}), three agentic coding harnesses driven by GPT-5.5, Claude-Sonnet-4.6, and DeepSeek-V4-Pro, respectively, and CodeArts \cite{codearts}, Huawei's official IDE plugin for HarmonyOS development, which offers AI-assisted code completion and generation with dedicated ArkTS and ArkUI support and is configured with GLM-5.2, the strongest model available within it. 
All models are accessed through their official APIs, and the harnesses and the IDE plugin run in their default configurations; the setups evaluated in each RQ are enumerated at the beginning of the corresponding subsection. % The IDE plugin joins only RQ1: as an interactive completion assistant, it exposes no scriptable interface through which the RQ2 strategies could be applied under the same protocol. 
% Qwen3.6-27B serves as the lightweight backbone because it is natively multimodal, and because it is among the strongest open-weight models in its class: its failures under direct translation and generic information thus upper-bound those of the entire lightweight class.
% via OpenAI's API, Kimi-K2-Turbo via Moonshot AI's API, GLM-5 via Zhipu AI's API, and Deepseek-v3.2 via DeepSeek's API. 
During the inference on all experiments, we set the temperature \(t=0\) and sampling number \(n=1\) to obtain deterministic outputs, minimizing randomness and ensuring reproducibility. Every generation and repair call issued by our scripts is capped at 16,384 output tokens, and the thinking mode is additionally disabled. All iterative-repair loops run within the five-round budget, whose sufficiency is empirically verified in Section~\ref{RQ3}, and all other hyperparameters are kept at their default values. For the construction of UI trees and ArkUI skeletons, we utilize tree-sitter-kotlin (v1.1.0) and tree-sitter-swift (v0.0.1) to parse the source code into ASTs. All prompts used in the evaluation are publicly available in our repository \cite{ArkTrans:online}.
% The prompts used for all experiments, including the baseline and our method, are available in our repository\cite{ArkTrans:online}. 
We conduct our experiments on Android 16 (API 36), iOS 18.6, and HarmonyOS 6.0.2. To ensure a consistent screen size across platforms, we use the iPhone SE (3rd generation) simulator for iOS and configure equivalent screen dimensions for Android and HarmonyOS.
% and (using an iPhone SE (3rd generation) simulator to match the target resolution)
The screen resolution is uniformly set to \(720\times1280\) at 320 DPI across all platforms, with the iOS simulator configured to the closest available dimensions (\(750\times1334\)) while maintaining the same physical density. 

\section{Evaluation Results}
% This section discusses the experimental results of \textit{ArkTrans}.

% Direct translation baseline （multi-model + agent frameworks）--0-35% CSR
\subsection{RQ1: Capability of Direct Translation }
\label{RQ1}

To measure how far model capability alone carries declarative UI migration, we test zero-shot direct translation on nine setups: five LLMs called directly, covering flagship and lightweight models (\textit{i.e.}, GPT-5.5 \cite{openai}, Claude-Sonnet-4.6 \cite{anthropic}, DeepSeek-V4-Pro \cite{deepseek}, GLM-5.2 \cite{zhipuai2024glm5}, and the lightweight model Qwen3.6-27B \cite{qwen}), three agentic coding harnesses (\textit{i.e.}, Codex driven by GPT-5.5, Claude Code driven by Claude-Sonnet-4.6, and the DeepSeek Harness (DSH) driven by DeepSeek-V4-Pro), and CodeArts \cite{codearts}, Huawei's official IDE plugin for HarmonyOS development, which offers AI-assisted code completion and generation with dedicated ArkTS and ArkUI support and is configured with GLM-5.2, the strongest model available within it. % We include this plugin as the platform-native baseline, since it is the assistant a HarmonyOS developer would actually use; it joins only RQ1, because as an interactive completion assistant it exposes no scriptable interface through which the RQ2 strategies could be applied under the same protocol. 
All setups receive the same zero-shot prompt, with no auxiliary information, and each of the 200 benchmark files (100 KJC + 100 SwiftUI) is translated independently.

\begin{table*}[htbp]

\centering
\setlength{\abovecaptionskip}{0.1cm}
\caption{Zero-shot translation performance of the nine setups}
\label{tab:rq1_direct}
\renewcommand{\arraystretch}{0.92}%
\resizebox{\linewidth}{!}{
\begin{tabular}{@{}lllccccccc@{}}
\toprule
\multirow{2}{*}{\textbf{Sou. PL}} & \multirow{2}{*}{\textbf{Category}} & \multirow{2}{*}{\textbf{Setup}} & \multirow{2}{*}{\textbf{CSR (\%)}} & \multicolumn{2}{c}{\textbf{Visual Fidelity (G)}} & \multicolumn{4}{c}{\textbf{Visual Fidelity (L)}} \\ \cmidrule(lr){5-6} \cmidrule(l){7-10}
 &  &  &  & \textbf{CH} & \textbf{CLIP} & \textbf{Pos} & \textbf{Size} & \textbf{Color} & \textbf{Text} \\ \midrule
\multirow{9}{*}{KJC}
 & \multirow{5}{*}{LLM-only}
   & GPT-5.5 & 29.00 & 20.73 & 25.85 & 25.82 & 26.20 & 22.41 & 28.44 \\
 &  & Claude-Sonnet-4.6 & 26.00 & 17.20 & 23.35 & 22.79 & 24.34 & 20.34 & 25.47 \\
 &  & DeepSeek-V4-Pro & 14.00 & 10.69 & 13.43 & 11.56 & 11.13 & 10.26 & 13.60 \\
 &  & GLM-5.2 & 5.00 & 1.41 & 2.91 & 2.49 & 2.44 & 2.33 & 2.87 \\
 &  & Qwen3.6-27B & 0.00 & 0.00 & 0.00 & 0.00 & 0.00 & 0.00 & 0.00 \\ \cmidrule(l){2-10}
 & \multirow{3}{*}{Agent harness}
   & Codex (GPT-5.5) & \textbf{35.00} & \textbf{21.59} & \textbf{31.37} & \textbf{27.91} & \textbf{29.33} & \textbf{24.74} & \textbf{34.30} \\
 &  & Claude Code (Sonnet-4.6) & 12.00 & 7.40 & 10.70 & 9.40 & 9.68 & 8.07 & 10.73 \\
 &  & DSH (DeepSeek-V4-Pro) & 9.00 & 8.04 & 7.82 & 8.16 & 7.82 & 7.10 & 8.86 \\ \cmidrule(l){2-10}
 & Huawei's IDE plugin& CodeArts (GLM-5.2) & 4.00 & 2.93 & 3.79 & 3.36 & 3.62 & 3.27 & 3.99 \\ \midrule
\multirow{9}{*}{SwiftUI}
 & \multirow{5}{*}{LLM-only}
   & GPT-5.5 & \textbf{25.00} & \textbf{18.25} & 22.27 & \textbf{21.30} & \textbf{22.20} & \textbf{19.88} & 24.33 \\
 &  & Claude-Sonnet-4.6 & \textbf{25.00} & 16.53 & \textbf{22.45} & 21.21 & 21.53 & 18.51 & \textbf{24.63} \\
 &  & DeepSeek-V4-Pro & 4.00 & 2.54 & 3.92 & 3.18 & 3.00 & 2.74 & 3.86 \\
 &  & GLM-5.2 & 1.00 & 0.00 & 0.93 & 0.61 & 0.64 & 0.27 & 0.91 \\
 &  & Qwen3.6-27B & 0.00 & 0.00 & 0.00 & 0.00 & 0.00 & 0.00 & 0.00 \\ \cmidrule(l){2-10}
 & \multirow{3}{*}{Agent harness}
   & Codex (GPT-5.5) & 19.00 & 10.14 & 16.82 & 15.59 & 15.93 & 13.60 & 18.70 \\
 &  & Claude Code (Sonnet-4.6) & 3.00 & 1.32 & 2.90 & 2.55 & 2.87 & 2.23 & 2.94 \\
 &  & DSH (DeepSeek-V4-Pro) & 0.00 & 0.00 & 0.00 & 0.00 & 0.00 & 0.00 & 0.00 \\ \cmidrule(l){2-10}
 & Huawei's IDE plugin& CodeArts (GLM-5.2) & 6.00 & 4.64 & 5.70 & 4.82 & 5.16 & 4.71 & 5.87 \\ \bottomrule
\end{tabular}
}
\par\vspace{2pt}
% \footnotesize
% \raggedright
% \noindent\textit{Note:} \textbf{G} denotes Global Metrics, \textbf{L} denotes Local Metrics.
% \normalsize
\begin{tablenotes}
    \footnotesize
    \item \textit{Note:} \textbf{G} denotes Global Metrics, \textbf{L} denotes Local Metrics.
\end{tablenotes}
\footnotesize
\raggedright

\end{table*}

% \begin{table}[htbp]
% \centering
% \setlength{\abovecaptionskip}{0.1cm}
% \caption{Direct translation performance of different entries}
% \label{tab:rq1_direct}
% \footnotesize
% \begin{tabular}{@{}llcc@{}}
% \toprule
% \multirow{2}{*}{\textbf{Category}} & \multirow{2}{*}{\textbf{Entry}} & \textbf{KJC} & \textbf{SwiftUI} \\
%  &  & \textbf{CSR (\%)} & \textbf{CSR (\%)} \\ \midrule
% \multirow{5}{*}{LLM-only}
%  & GPT-5.5 & 29.00 & 25.00 \\
%  & Claude-Sonnet-4.6 & 26.00 & 25.00 \\
%  & DeepSeek-V4-Pro & 14.00 & 4.00 \\
%  & GLM-5.2 & 5.00 & 1.00 \\
%  & Qwen3.6-27B & 0.00 & 0.00 \\ \midrule
% \multirow{2}{*}{Agent harness}
%  & Codex (GPT-5.5) & 35.00 & 19.00 \\
%  & DSH (DeepSeek-V4-Pro) & 9.00 & 0.00 \\ \midrule
% Huawei's IDE plugin& CodeArts (GLM-5.2) & 4.00 & 6.00 \\ \bottomrule
% \end{tabular}
% \par\vspace{2pt}
% \footnotesize
% \flushleft
% \normalsize
% \end{table}

Table~\ref{tab:rq1_direct} summarizes the results. Across all nine setups, CSR spans 0.00\%--35.00\%: not a single setup reliably produces compilable ArkUI code. Among LLM-only setups, the best result is 29.00\% (GPT-5.5 on KJC), and on SwiftUI the best is a 25.00\% tie between GPT-5.5 and Claude-Sonnet-4.6; the lightweight model Qwen3.6-27B fails completely, producing no compilable ArkUI code in either translation direction.

One might attribute the failure to the simplicity of a direct LLM call, yet the agentic setups tell otherwise. Codex improves over the direct call of its own backbone by only 6.00 percentage points on KJC, from 29.00\% to 35.00\%, and falls behind it on SwiftUI, where the backbone reaches 25.00\% but Codex only 19.00\%. More strikingly, the DSH harness scores below its own backbone, 9.00\% versus 14.00\% on KJC and 0.00\% versus 4.00\% on SwiftUI, indicating that unconstrained agent autonomy can introduce noise faster than it removes errors. The vendor's official IDE plugin, CodeArts, stays at 4.00\% on KJC and 6.00\% on SwiftUI: even the vendor's own tooling fails, which rules out tool immaturity as an explanation. These results confirm that the obstacle is not the lack of an execution loop or IDE integration, but the paradigm gap itself: without dedicated guidance, neither LLMs nor agent frameworks can consistently emit error-free ArkUI code from prior knowledge alone.

\begin{boxK}
\small \faIcon{pencil-alt} \textbf{Finding 1:}
Direct translation caps at 35.00\% CSR across all nine setups, and agentic harnesses and the official IDE plugin rarely outperform the direct calls of their own backbones. The obstacle is the paradigm gap itself rather than the lack of tooling, and the lightweight model fails completely.
\end{boxK}

\subsection{RQ2: Performance of Common Strategies}
\label{RQ2}

Table~\ref{tab:rq2_remedies} reports the evaluation of the two common strategies on the same benchmark, with the zero-shot rows of RQ1 carried over as baselines. CodeArts is not evaluated as it is an IDE plugin without a scriptable interface. The reference image requires multimodal input. Codex and Claude Code run both strategies on their multimodal backbones, whereas GLM-5.2 and DeepSeek-V4-Pro are text-only models, so GLM-5.2 and DSH with DeepSeek-V4-Pro are evaluated with iterative repair only.

\textbf{Reference images ground visual intent but cannot correct syntax.}
Attaching the rendered screenshot of the source UI yields the largest
single-shot gain observed in this study: GPT-5.5 rises from 29.00\%/25.00\% to 61.00\%/54.00\% CSR, and Claude-Sonnet-4.6 from 26.00\%/25.00\% to 75.00\%/64.00\%. 
Visual fidelity follows suit: for Claude-Sonnet-4.6 on KJC, CH climbs from 17.20 to 55.48, Pos from 22.79 to 66.02, and Text from 25.47 to 73.12, confirming that the image effectively channels layout and styling intent into the translation. The same effect appears for the agent harnesses as well: given the reference image, Claude Code improves from 12.00\% to 14.00\% CSR on KJC and from 3.00\% to 9.00\% on SwiftUI, and Codex from 35.00\% to 43.00\% and from 19.00\% to 38.00\%. 

Nevertheless, the image brings Qwen3.6-27B almost no gain, 2.00\% CSR on KJC and 0.00\% on SwiftUI: visual grounding cannot repair broken ArkTS syntax.

\begin{boxK}
\small \faIcon{pencil-alt} \textbf{Finding 2:}
A single reference image lifts the single-call CSR of capable LLMs by 29--49 percentage points and improves every visual metric in step.
\end{boxK}

\begin{table*}[htbp]

\centering
\setlength{\abovecaptionskip}{0.1cm}
\caption{Translation performance with common strategies}% \caption{Effect of reference images and compiler iteration}
\label{tab:rq2_remedies}
\renewcommand{\arraystretch}{0.92}%
\small
\resizebox{\linewidth}{!}{
\begin{tabular}{@{}llllccccccc@{}}
\toprule
\multirow{2}{*}{\textbf{Sou. PL}} & \multirow{2}{*}{\textbf{Category}} & \multirow{2}{*}{\textbf{Setup}} & \multirow{2}{*}{\textbf{Method}} & \multirow{2}{*}{\textbf{CSR (\%)}} & \multicolumn{2}{c}{\textbf{Visual Fidelity (G)}} & \multicolumn{4}{c}{\textbf{Visual Fidelity (L)}} \\ \cmidrule(lr){6-7} \cmidrule(l){8-11}
 &  &  &  &  & \textbf{CH} & \textbf{CLIP} & \textbf{Pos} & \textbf{Size} & \textbf{Color} & \textbf{Text} \\ \midrule
\multirow{22}{*}{KJC}
 & \multirow{12}{*}{LLM-only}
 & GPT-5.5 & Direct (RQ1) & 29.00 & 20.73 & 25.85 & 25.82 & 26.20 & 22.41 & 28.44 \\
 &  &  & + Image & \cellcolor{gray!25}61.00 & \cellcolor{gray!25}43.22 & \cellcolor{gray!25}53.87 & \cellcolor{gray!25}53.03 & \cellcolor{gray!25}55.52 & \cellcolor{gray!25}47.11 & \cellcolor{gray!25}59.52 \\
 &  & Claude-Sonnet-4.6 & Direct (RQ1) & 26.00 & 17.20 & 23.35 & 22.79 & 24.34 & 20.34 & 25.47 \\
 &  &  & + Image & 75.00 & 55.48 & 67.11 & 66.02 & 66.79 & 57.61 & 73.12 \\
 &  &  & + Iterative Repair & 85.00 & 62.88 & 75.07 & 70.92 & 73.13 & 62.40 & 83.14 \\
 &  &  & + Image + Iterative Repair & \cellcolor{gray!25}94.00 & \cellcolor{gray!25}69.28 & \cellcolor{gray!25}83.16 & \cellcolor{gray!25}80.79 & \cellcolor{gray!25}83.07 & \cellcolor{gray!25}71.09 & \cellcolor{gray!25}92.05 \\
 &  & GLM-5.2 & Direct (RQ1) & 5.00 & 1.41 & 2.91 & 2.49 & 2.44 & 2.33 & 2.87 \\
 &  &  & + Iterative Repair & \cellcolor{gray!25}55.00 & \cellcolor{gray!25}36.70 & \cellcolor{gray!25}47.98 & \cellcolor{gray!25}44.93 & \cellcolor{gray!25}43.79 & \cellcolor{gray!25}39.45 & \cellcolor{gray!25}53.32 \\
 &  & Qwen3.6-27B & Direct (RQ1) & 0.00 & 0.00 & 0.00 & 0.00 & 0.00 & 0.00 & 0.00 \\
 &  &  & + Image & 2.00 & 1.52 & 1.96 & 1.64 & 1.73 & 1.40 & 1.94 \\
 &  &  & + Iterative Repair & 20.00 & 8.96 & 11.32 & 9.09 & 9.63 & 8.02 & 10.57 \\
 &  &  & + Image + Iterative Repair & \cellcolor{gray!25}26.00 & \cellcolor{gray!25}18.32 & \cellcolor{gray!25}23.42 & \cellcolor{gray!25}19.32 & \cellcolor{gray!25}18.41 & \cellcolor{gray!25}15.79 & \cellcolor{gray!25}22.83 \\ \cmidrule(l){2-11}
 & \multirow{10}{*}{Agent harness}
 & Codex (GPT-5.5) & Harness direct (RQ1) & 35.00 & 21.59 & 31.37 & 27.91 & 29.33 & 24.74 & 34.30 \\
 &  &  & + Image & 43.00 & 29.10 & 38.76 & 31.12 & 32.17 & 27.88 & 38.89 \\
 &  &  & + Iterative Repair & \cellcolor{gray!25}\textbf{100.00} & \cellcolor{gray!25}74.46 & \cellcolor{gray!25}88.69 & \cellcolor{gray!25}82.69 & \cellcolor{gray!25}82.63 & \cellcolor{gray!25}73.46 & \cellcolor{gray!25}97.42 \\
 &  &  & + Image + Iterative Repair & 98.00 & 68.48 & 88.51 & 67.95 & 69.22 & 62.27 & 88.17 \\
 &  & Claude Code (Sonnet-4.6) & Harness direct (RQ1) & 12.00 & 7.40 & 10.70 & 9.40 & 9.68 & 8.07 & 10.73 \\
 &  &  & + Image & 14.00 & 6.94 & 12.26 & 10.21 & 10.75 & 8.98 & 12.22 \\
 &  &  & + Iterative Repair & \cellcolor{gray!25}\textbf{100.00} & 69.13 & \cellcolor{gray!25}\textbf{93.92} & \cellcolor{gray!25}82.23 & \cellcolor{gray!25}82.82 & 70.74 & \cellcolor{gray!25}95.46 \\
 &  &  & + Image + Iterative Repair & \cellcolor{gray!25}\textbf{100.00} & \cellcolor{gray!25}71.60 & 93.61 & 78.71 & 78.62 & \cellcolor{gray!25}71.15 & 93.14 \\
 &  & DSH (DeepSeek-V4-Pro) & Harness direct (RQ1) & 9.00 & 8.04 & 7.82 & 8.16 & 7.82 & 7.10 & 8.86 \\
 &  &  & + Iterative Repair & \cellcolor{gray!25}\textbf{100.00} & \cellcolor{gray!25}\textbf{75.99} & \cellcolor{gray!25}88.36 & \cellcolor{gray!25}\textbf{85.51} & \cellcolor{gray!25}\textbf{88.64} & \cellcolor{gray!25}\textbf{75.78} & \cellcolor{gray!25}\textbf{97.65} \\ \midrule
\multirow{22}{*}{SwiftUI}
 & \multirow{12}{*}{LLM-only}
 & GPT-5.5 & Direct (RQ1) & 25.00 & 18.25 & 22.27 & 21.30 & 22.20 & 19.88 & 24.33 \\
 &  &  & + Image & \cellcolor{gray!25}54.00 & \cellcolor{gray!25}39.68 & \cellcolor{gray!25}48.02 & \cellcolor{gray!25}46.26 & \cellcolor{gray!25}48.46 & \cellcolor{gray!25}41.75 & \cellcolor{gray!25}52.56 \\
 &  & Claude-Sonnet-4.6 & Direct (RQ1) & 25.00 & 16.53 & 22.45 & 21.21 & 21.53 & 18.51 & 24.63 \\
 &  &  & + Image & 64.00 & 50.35 & 56.66 & 55.00 & 56.42 & 48.71 & 62.53 \\
 &  &  & + Iterative Repair & 95.00 & 70.11 & 84.10 & 80.01 & 81.78 & 71.54 & 92.58 \\
 &  &  & + Image + Iterative Repair & \cellcolor{gray!25}99.00 & \cellcolor{gray!25}73.79 & \cellcolor{gray!25}87.69 & \cellcolor{gray!25}82.81 & \cellcolor{gray!25}\textbf{85.88} & \cellcolor{gray!25}73.79 & \cellcolor{gray!25}96.48 \\
 &  & GLM-5.2 & Direct (RQ1) & 1.00 & 0.00 & 0.93 & 0.61 & 0.64 & 0.27 & 0.91 \\
 &  &  & + Iterative Repair & \cellcolor{gray!25}68.00 & \cellcolor{gray!25}51.48 & \cellcolor{gray!25}59.95 & \cellcolor{gray!25}54.85 & \cellcolor{gray!25}53.78 & \cellcolor{gray!25}49.72 & \cellcolor{gray!25}66.11 \\
 &  & Qwen3.6-27B & Direct (RQ1) & 0.00 & 0.00 & 0.00 & 0.00 & 0.00 & 0.00 & 0.00 \\
 &  &  & + Image & 0.00 & 0.00 & 0.00 & 0.00 & 0.00 & 0.00 & 0.00 \\
 &  &  & + Iterative Repair & 6.00 & 3.93 & 3.89 & 3.79 & 3.93 & 3.46 & 3.97 \\
 &  &  & + Image + Iterative Repair & \cellcolor{gray!25}13.00 & \cellcolor{gray!25}7.02 & \cellcolor{gray!25}12.26 & \cellcolor{gray!25}9.83 & \cellcolor{gray!25}10.52 & \cellcolor{gray!25}8.74 & \cellcolor{gray!25}11.73 \\ \cmidrule(l){2-11}
 & \multirow{10}{*}{Agent harness}
   & Codex (GPT-5.5) & Harness direct (RQ1) & 19.00 & 10.14 & 16.82 & 15.59 & 15.93 & 13.60 & 18.70 \\
 &  &  & + Image & 38.00 & 28.47 & 35.44 & 27.93 & 29.49 & 26.01 & 35.58 \\
 &  &  & + Iterative Repair & \cellcolor{gray!25}\textbf{100.00} & \cellcolor{gray!25}73.29 & 88.25 & \cellcolor{gray!25}\textbf{84.62} & \cellcolor{gray!25}85.12 & \cellcolor{gray!25}\textbf{74.60} & \cellcolor{gray!25}\textbf{97.31} \\
 &  &  & + Image + Iterative Repair & 98.00 & 69.31 & \cellcolor{gray!25}90.46 & 71.34 & 74.91 & 66.49 & 90.16 \\
 &  & Claude Code (Sonnet-4.6) & Harness direct (RQ1) & 3.00 & 1.32 & 2.90 & 2.55 & 2.87 & 2.23 & 2.94 \\
 &  &  & + Image & 9.00 & 6.20 & 8.54 & 8.25 & 8.10 & 7.16 & 8.92 \\
 &  &  & + Iterative Repair & \cellcolor{gray!25}\textbf{100.00} & \cellcolor{gray!25}71.41 & \cellcolor{gray!25}\textbf{92.07} & 78.02 & 76.77 & 69.33 & \cellcolor{gray!25}93.99 \\
 &  &  & + Image + Iterative Repair & 98.00 & 68.75 & 90.82 & \cellcolor{gray!25}78.14 & \cellcolor{gray!25}79.44 & \cellcolor{gray!25}69.77 & 93.14 \\
 &  & DSH (DeepSeek-V4-Pro) & Harness direct (RQ1) & 0.00 & 0.00 & 0.00 & 0.00 & 0.00 & 0.00 & 0.00 \\
 &  &  & + Iterative Repair & \cellcolor{gray!25}98.00 & \cellcolor{gray!25}\textbf{74.20} & \cellcolor{gray!25}88.22 & \cellcolor{gray!25}81.16 & \cellcolor{gray!25}84.08 & \cellcolor{gray!25}72.58 & \cellcolor{gray!25}94.98 \\ \bottomrule

\end{tabular}
}
\begin{tablenotes}
    \footnotesize
    \item \textit{Note:} \textbf{G} denotes Global Metrics, \textbf{L} denotes Local Metrics. Shaded cells mark the best result of each setup in each column. Bold marks the best across all setups.
\end{tablenotes}
\end{table*}

\textbf{Iterative repair rescues capable models.}
Feeding real hvigor diagnostics back within a five-round iteration budget transforms the capable models: Claude-Sonnet-4.6 reaches 85.00\%/95.00\% CSR, all three harnesses converge to 98.00--100.00\%, and even the weakest commercial model GLM-5.2 jumps from 5.00\%/1.00\% to 55.00\%/68.00\%. Combining both strategies pushes Claude-Sonnet-4.6 to 94.00\%/99.00\% CSR, with CH of 69.28/73.79 and Text of 92.05/96.48.
% Interestingly, iteration also inverts the direction gap: SwiftUI is seemingly the harder direction in RQ1, yet after iteration capable models match or exceed their KJC results (\textit{e.g.}, 95.00\% vs. 85.00\% for Claude, 68.00\% vs. 55.00\% for GLM-5.2), suggesting that SwiftUI's residual errors are predominantly compiler-detectable and thus repairable, whereas part of KJC's errors survive the loop.

\begin{boxK}
\small \faIcon{pencil-alt} \textbf{Finding 3:}
Iterative repair performs remarkably well: capable setups reach 94--100\% CSR within the maximum iteration budget, essentially solving executability where model capability suffices.
\end{boxK}

\textbf{Common strategies are capability-gated.}
Under the same strategies, Qwen3.6-27B improves only from 0.00\% to 26.00\% CSR on KJC and 13.00\% on SwiftUI, still failing to compile about three of every four KJC files and nearly nine of every ten SwiftUI files; each single strategy is weaker still, with compiler iterative repair alone reaching 20.00\%/6.00\% and the reference image alone almost nothing. Its visual fidelity under the combined strategy remains marginal, with CH at 18.32/7.02. Worse, the iteration loop burns 21,758 tokens per KJC file and 26,588 per SwiftUI file on average without converging. Such failures do not come from a weak pick. As a natively multimodal LLM, it can take the reference image as input, and it is among the strongest open-weight models in its class; its failure under both strategies thus upper-bounds what the entire lightweight class can gain from generic information.

The payoff is in fact monotone in base capability: the stronger the zero-shot baseline, the more the strategies deliver, from near-perfect restoration for the capable models, down to a marginal 26.00\%/13.00\% for Qwen3.6-27B. We attribute this to the nature of the two strategies: both are generic information. The reference image specifies \textit{what} the UI should look like, and the compiler pinpoints \textit{where} the code goes wrong, but neither conveys \textit{how} correct ArkTS should be written. A model with approximate ArkUI knowledge can act on such a signal and repair itself; a model without it can only re-sample its own failure mode, round after round. RQ3 verifies this hypothesis by dissecting the iteration traces of Qwen3.6-27B, and the observation directly shapes our design principle in RQ4: when feedback cannot substitute for knowledge, supply structure and knowledge rather than more feedback.

\begin{boxK}
\small \faIcon{pencil-alt} \textbf{Finding 4:}
Common strategies are capability-gated: their payoff grows monotonically with base capability, and Qwen3.6-27B reaches only 26.00\%/13.00\% CSR even with both strategies. Generic feedback tells a model \textit{what} and \textit{where}, but not \textit{how} to write correct ArkTS.
\end{boxK}

\subsection{RQ3: Why the Common Strategies Fail on the Lightweight Model}
\label{RQ3}

This section examines the errors that the lightweight model (Qwen3.6-27B) made under all four configurations of RQ2, aiming to identify why none of the configurations exceeds 26.00\% CSR on KJC and 13.00\% on SwiftUI. We characterize the recurring failure patterns through compiler diagnostics and examine how these errors evolve under reference images and iterative repair, thereby identifying the limitations of common strategies.
Following established thematic-analysis and open-coding procedures \cite{cruzes2011recommended}, we analyze 29,593 compiler diagnostics collected from the generated code and repair traces across all four configurations. We draw a stratified sample of 3,102 diagnostics by error code at a 95\% confidence level with a $\pm$5\% margin of error, and three ArkTS experts manually open-code it into 267 signatures consolidated into a two-level codebook in 120 person-hours, achieving substantial inter-coder agreement (Cohen's $\kappa = 0.84$). We then derive a regular-expression pattern for each error family and apply these patterns to generalize the codebook to all 29,593 occurrences. The complete codebook, including representative code excerpts and compiler diagnostics, is provided in our replication package \cite{ArkTrans:online}.

The analysis identifies five major error families and 13 subfamilies, covering structural defects, nonexistent APIs, invalid ArkTS usage, misplaced UI code, and missing resources. Across the full corpus, nonexistent APIs (B, 32.0\%) and wrong usage (C, 24.8\%) are the most frequent families, followed by broken structure (A, 24.1\%), missing resources (E, 10.0\%), and misplaced UI code (D, 8.4\%). The four largest subfamilies (B-1, A-2, B-2, and C-1) jointly account for 17,918 diagnostics (60.5\%), indicating that failures are concentrated in a small set of recurring patterns. Table~\ref{tab:rq3_error_types} reports the final-state diagnostics of each configuration.

%, a subset of this total. 
The four configurations are abbreviated S1--S4 (direct, + Image, + Iterative Repair, +Image + Iterative Repair), and comparisons between them are written as S1 $\rightarrow$ S2.
The overall distribution shows that the two directions fail in categorically different ways at generation time: \textbf{broken structure (A)} accounts for the vast majority on KJC at 62.6\%, whereas \textbf{nonexistent APIs (B)} and \textbf{wrong usage (C)} jointly dominate SwiftUI at 36.6\% and 35.8\%, 72.4\% combined. In the final state of every iteration, regardless of configuration, \textbf{B} remains the largest family at 31.6--45.0\%, and \textbf{missing resources (E)} becomes the runner-up in the combined-strategy finals at 31.6\% and 24.6\%; \textbf{misplaced UI code (D)} and miscellaneous errors stay marginal throughout. Below, we introduce each error in detail.

\begin{table}[t]

\scriptsize
\setlength\tabcolsep{3pt}
\renewcommand{\arraystretch}{0.92}%
\caption{Error categories vs.\ translation directions and strategies}
\vspace{-1.5em}
\label{tab:rq3_error_types}
\begin{threeparttable}
\resizebox{\linewidth}{!}{%
\begin{tabular}{lc>{\columncolor{red!32}}c>{\columncolor{red!22}}c>{\columncolor{red!22}}c>{\columncolor{red!22}}c>{\columncolor{red!22}}c>{\columncolor{yellow!32}}c>{\columncolor{yellow!22}}c>{\columncolor{yellow!22}}c>{\columncolor{blue!28}}c>{\columncolor{blue!16}}c>{\columncolor{blue!16}}c>{\columncolor{blue!16}}c>{\columncolor{green!20}}c>{\columncolor{green!12}}c>{\columncolor{green!12}}c>{\columncolor{green!12}}c>{\columncolor{brown!16}}c>{\columncolor{gray!12}}c}
\toprule
\textbf{Categories} & \textbf{N} & \textbf{A} & \textbf{A-1} & \textbf{A-2} & \textbf{A-3} & \textbf{A-4} & \textbf{B} & \textbf{B-1} & \textbf{B-2} & \textbf{C} & \textbf{C-1} & \textbf{C-2} & \textbf{C-3} & \textbf{D} & \textbf{D-1} & \textbf{D-2} & \textbf{D-3} & \textbf{E} & \textbf{Oth.} \\
\midrule
\multicolumn{20}{l}{\textbf{S1: Direct}} \\ \midrule
\hspace{1.5em}KJC $\rightarrow$ ArkUI & 5,830 & \textbf{62.6} & 2.6 & \textbf{58.9} & 0.0 & 1.0 & 15.4 & 6.4 & 9.0 & 14.2 & 6.1 & 0.8 & 7.3 & 4.9 & 4.6 & 0.2 & 0.1 & 1.6 & \textbf{1.4} \\
\hspace{1.5em}SwiftUI $\rightarrow$ ArkUI & 3,531 & 13.5 & \textbf{9.3} & 1.1 & \textbf{0.3} & \textbf{2.8} & \textbf{36.6} & \textbf{19.8} & \textbf{16.8} & \textbf{35.8} & \textbf{22.7} & \textbf{1.2} & \textbf{11.9} & \textbf{9.8} & \textbf{9.1} & \textbf{0.5} & \textbf{0.2} & \textbf{3.3} & 0.9 \\
\midrule
\multicolumn{20}{l}{\textbf{S2: + Image}} \\ \midrule
\hspace{1.5em}KJC $\rightarrow$ ArkUI & 2,018 & 10.2 & \textbf{8.5} & 1.2 & 0.5 & 0.0 & \textbf{40.8} & 13.3 & \textbf{27.5} & 23.7 & 9.3 & \textbf{2.0} & \textbf{12.4} & 4.9 & 4.4 & \textbf{0.4} & 0.1 & \textbf{7.0} & \textbf{13.3} \\
\hspace{1.5em}SwiftUI $\rightarrow$ ArkUI & 3,693 & \textbf{21.9} & 8.2 & \textbf{10.8} & \textbf{2.9} & 0.0 & 32.2 & \textbf{15.1} & 17.1 & \textbf{26.4} & \textbf{13.7} & 1.2 & 11.5 & \textbf{8.7} & \textbf{8.4} & 0.2 & 0.1 & 4.2 & 6.6 \\
\midrule
\multicolumn{20}{l}{\textbf{S3: + Iterative Repair}} \\ \midrule
\hspace{1.5em}KJC $\rightarrow$ ArkUI & 736 & 4.5 & \textbf{4.5} & 0.0 & 0.0 & 0.0 & \textbf{45.0} & 26.9 & \textbf{18.1} & 24.9 & 9.8 & \textbf{5.4} & 9.6 & \textbf{11.5} & \textbf{9.0} & 1.1 & 1.5 & 13.5 & \textbf{0.7} \\
\hspace{1.5em}SwiftUI $\rightarrow$ ArkUI & 914 & \textbf{9.4} & 3.7 & \textbf{4.3} & \textbf{1.4} & 0.0 & 39.3 & \textbf{28.0} & 11.3 & \textbf{27.0} & \textbf{14.2} & 2.5 & \textbf{10.3} & 7.2 & 4.0 & \textbf{1.6} & 1.5 & \textbf{17.0} & 0.1 \\
\midrule
\multicolumn{20}{l}{\textbf{S4: + Image + Iterative Repair}} \\ \midrule
\hspace{1.5em}KJC $\rightarrow$ ArkUI & 455 & 15.2 & \textbf{4.4} & 8.1 & 2.6 & 0.0 & \textbf{31.9} & 16.3 & \textbf{15.6} & 19.6 & 7.3 & \textbf{4.6} & \textbf{7.7} & 1.8 & 0.7 & 0.4 & 0.7 & \textbf{31.6} & 0.0 \\
\hspace{1.5em}SwiftUI $\rightarrow$ ArkUI & 817 & \textbf{17.1} & 2.7 & \textbf{10.4} & \textbf{4.0} & 0.0 & 31.6 & \textbf{18.0} & 13.6 & 19.6 & \textbf{13.1} & 1.8 & 4.7 & \textbf{6.9} & \textbf{5.5} & \textbf{0.5} & \textbf{0.9} & 24.6 & \textbf{0.2} \\
\bottomrule
\end{tabular}%
}
\end{threeparttable}

\end{table}

\begin{figure}[htbp]
  \centering
  \includegraphics[width=\textwidth]{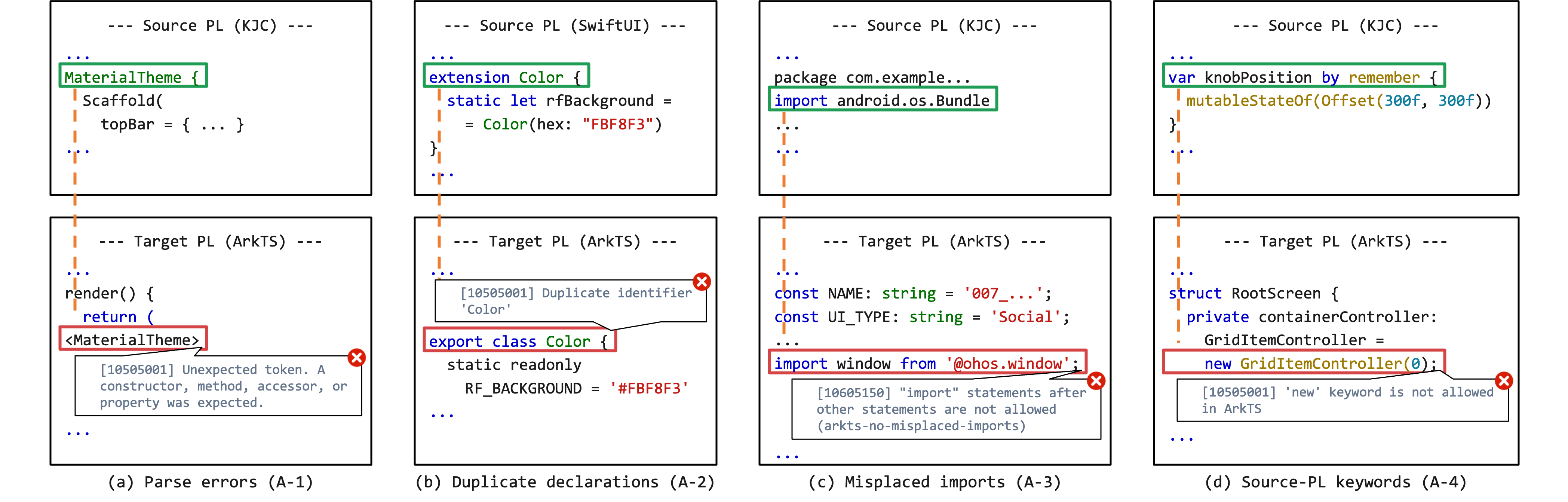}
  \caption{Representative errors of broken structure (A).}
  \label{fig:rq3_A}
\end{figure}

\textbf{Broken structure (A)} is the debris of unstructured whole-file generation. 
In the direct runs on KJC, duplicate declarations (A-2) account for 94.1\% of A, typically involving repeated component declarations or conflicting identifiers. For instance, one generated file defines \texttt{\detokenize{export class Color}}, which collides with ArkUI's built-in \texttt{Color} identifier and triggers a duplicate-identifier error (Figure~\ref{fig:rq3_A}(b)). Other structural failures include parse errors (A-1), such as JSX-like markup emitted in ArkTS files (Figure~\ref{fig:rq3_A}(a)); misplaced imports (A-3), where import statements appear after variable declarations (Figure~\ref{fig:rq3_A}(c)); and source-language keywords retained in ArkTS (A-4), such as \texttt{\detokenize{new GridItemController(0)}} (Figure~\ref{fig:rq3_A}(d)). On SwiftUI, the direct profile differs: parse errors (A-1) account for 68.9\% of A, while A as a whole represents only 13.5\% of all diagnostics. These differences indicate that structural failures are strongly dependent on the source translation direction.

% In the direct runs on KJC, it consists almost entirely of duplicate declarations (A-2, the same \texttt{@Component} struct emitted twice in one file), which account for 94.1\% of A, followed by parse errors (A-1) at 4.2\% and source-PL keywords (A-4) at 1.6\%, which are exclusively the Swift-style \texttt{new} pasted into constructor calls such as \texttt{new Column()}, with the remaining misplaced imports (A-3), i.e., \texttt{import} statements placed after other code, staying marginal throughout. On SwiftUI the direct profile inverts, with A-1 at 68.9\% and A-2 at 8.1\% of \textbf{A}: the file largely parses, and structure is not the disease.

\begin{figure}[htbp]
  \centering
  \includegraphics[width=\textwidth]{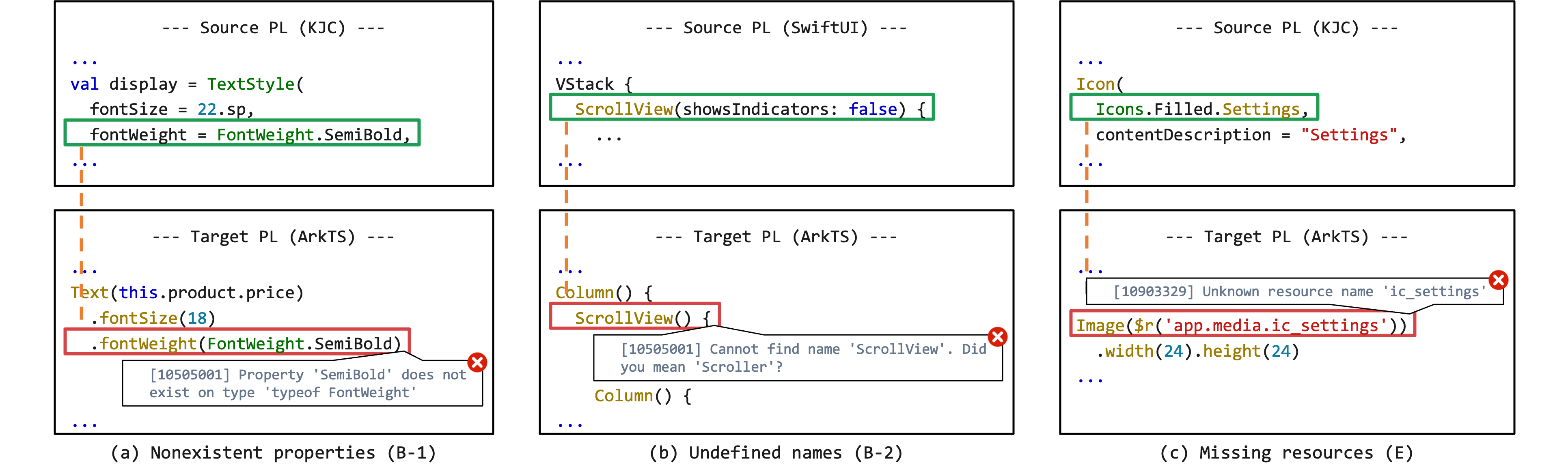}
  \caption{Representative errors of nonexistent APIs (B) and missing resources (E).}
  \label{fig:rq3_BE}
\end{figure}

\textbf{Nonexistent APIs (B)} arise when the model generates properties, methods, or modules that are unavailable in ArkUI. For nonexistent properties and calls (B-1), a representative example is \texttt{FontWeight.SemiBold} (Figure~\ref{fig:rq3_BE}(a)), which is rejected because \texttt{SemiBold} is not a member of the target \texttt{FontWeight} enum. 
Another case chains \texttt{.margin(...)} onto a method returning \texttt{void}, incorrectly treating it as a UI component. For undefined names and modules (B-2), the model directly carries over the SwiftUI component \texttt{ScrollView()} into ArkTS (Figure~\ref{fig:rq3_BE}(b)).

% Undefined names and modules (B-2) lead on KJC at 58.4\% of B, imports that resolve to nothing; nonexistent properties and calls (B-1) lead on SwiftUI at 54.1\%, \textit{e.g.}, \texttt{.scrollDirection(Axis.Vertical)} on a \texttt{Scroll} that has no such attribute, or accessing \texttt{InputType.Text} as a property. 
A diagnostic only says that such a name or attribute does not exist, never which ArkTS API should be used instead; consistently, B remains the largest error family in the final states of both repair configurations (S3 and S4), accounting for 31.6--45.0\% of diagnostics.

\begin{figure}[htbp]
  \centering
  \includegraphics[width=\textwidth]{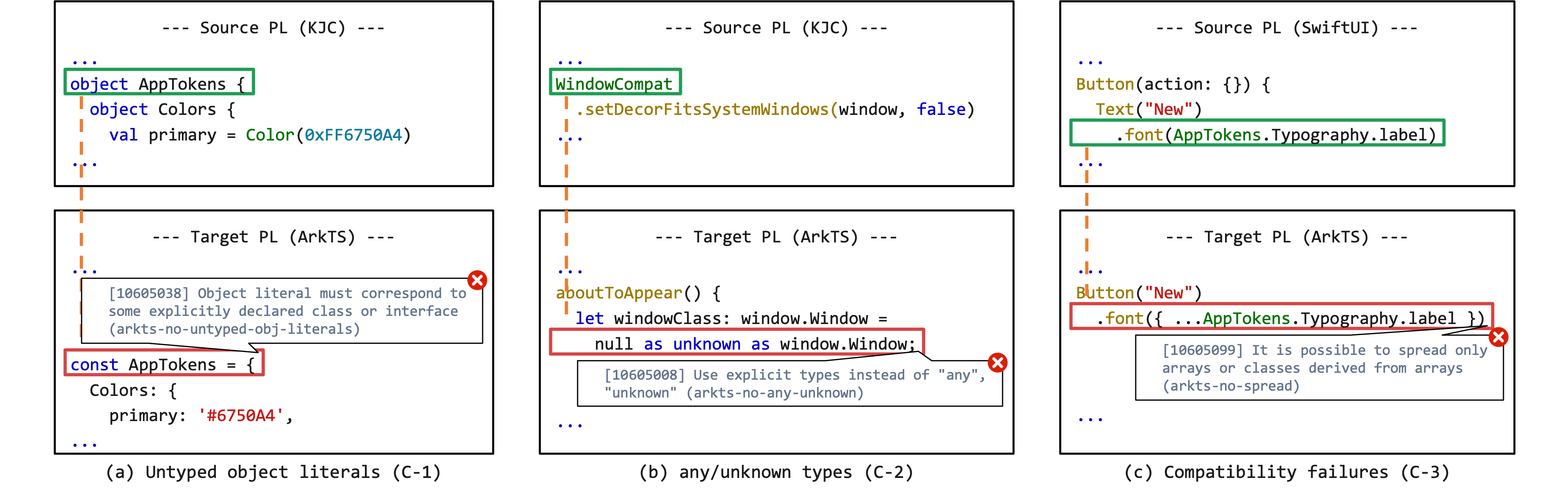}
  \caption{Representative errors of wrong usage (C).}
  \label{fig:rq3_C}
\end{figure}

\textbf{Wrong usage (C)} occurs when the generated code uses otherwise valid constructs in ways prohibited by ArkTS. The most frequent case is untyped object literals (C-1), where the model generates nested design-token objects such as \texttt{\detokenize{const AppTokens = {...}}} without explicitly declared classes or interfaces (Figure~\ref{fig:rq3_C}(a)). Such nested literals can trigger multiple \emph{arkts-no-untyped-obj-literals} diagnostics within a single file. % In the direct runs it takes two main forms: untyped object literals (C-1), which account for 63.4\% of C on SwiftUI, typically \texttt{\{text:~'...'\}} passed without a declared interface; and compatibility failures (C-3) at 33.2\%, whose typical case is passing a plain string where \texttt{TextInputOptions} is expected, alongside the \emph{arkts-*} language restrictions on spread and comma operators. The remaining explicit-type violations (C-2), i.e., \texttt{any}/\texttt{unknown} used where ArkTS demands explicit types, stay marginal.
Other violations include the use of \texttt{any}/\texttt{unknown} types (C-2), such as \texttt{\detokenize{as unknown as window.Window}} (Figure~\ref{fig:rq3_C}(b)), and compatibility failures (C-3), such as \texttt{\detokenize{...AppTokens.Typography.label}} in an object literal (Figure~\ref{fig:rq3_C}(c)), which violates \emph{arkts-no-spread}. The latter subfamily spans 53 distinct signatures, reflecting a variety of ArkTS-specific language restrictions. In the SwiftUI direct runs, untyped object literals alone account for 22.7\% of all errors, making incorrect ArkTS usage a major source of compilation failures.

\begin{figure}[htbp]
  \centering
  \includegraphics[width=\textwidth]{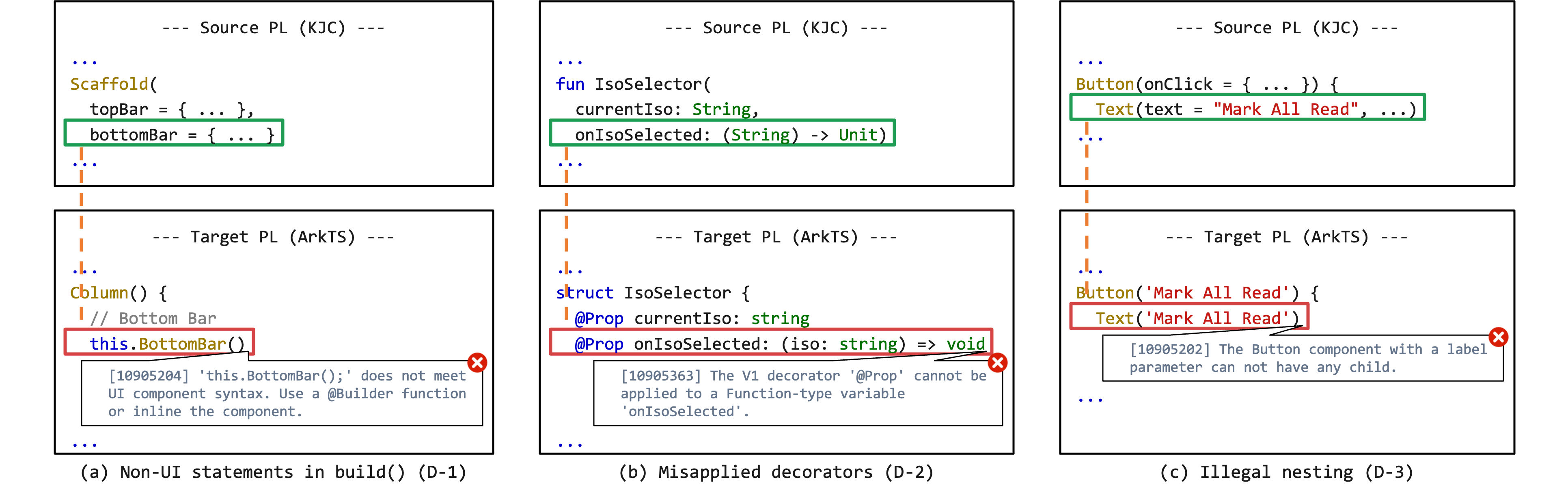}
  \caption{Representative errors of misplaced UI code (D).}
  \label{fig:rq3_D}
\end{figure}

\textbf{Misplaced UI code (D)} is small but persistent. It  refers to otherwise valid code placed in contexts that violate ArkUI's declarative UI constraints. The dominant subfamily is non-UI statements inside \texttt{build()} (D-1), such as invoking \texttt{this.BottomBar()} as an ordinary method (Figure~\ref{fig:rq3_D}(a)) or declaring local variables directly within a UI builder. The remaining cases involve misapplied decorators (D-2), such as \texttt{\detokenize{@Prop onIsoSelected: (iso: string) => void}} (Figure~\ref{fig:rq3_D}(b)), which applies \texttt{@Prop} to a function-typed variable, and illegal component nesting (D-3), such as placing a \texttt{Text} child inside a \texttt{Button} already initialized with a label (Figure~\ref{fig:rq3_D}(c)).
Although D represents a smaller proportion of errors than B and C, these cases expose ArkUI-specific composition constraints that cannot be inferred from the source UI hierarchy alone.

%, 1.8--11.5\% of all errors, and \textbf{non-UI statements inside \texttt{build()} (D-1)} account for about 70\% of D in the iteration finals, such as plain \texttt{let} declarations or \texttt{if} statements in the UI builder. The rest are misapplied decorators (D-2), e.g., a state decorator on a function-type variable, and illegal component nesting (D-3), e.g., children placed under a labeled \texttt{Button}, both marginal. 

\textbf{Missing resources (E)} references colors, strings, or media absent from the target project, \textit{e.g.}, \$r('app.media.ic\_settings') (Figure~\ref{fig:rq3_BE}(c)) for a default icon defined only in the source app. E is the only family whose absolute count grows under repair, from 94 to 144 errors on KJC and from 116 to 201 on SwiftUI while the totals collapse twelvefold on KJC and fourfold on SwiftUI. Thus far we have established what the lightweight model gets wrong; the next question is why none of the strategies exceeds 26.00\%/13.00\% CSR.

\textbf{The reference image transmits structure, not knowledge.} The reference image collapses broken structure on KJC (S1 $\rightarrow$ S2): duplicate declarations (A-2) fall from 58.9\% to 1.2\% of errors and the total drops by 65\%, from 5,830 to 2,018. The mechanism is visible in the composition: given a complete layout image, the model copies and tries to understand the structure it sees. However, that is all the image conveys. 
On the knowledge side, the gap is not reduced but exposed: in the S2 final state, nonexistent APIs (B) rise to 40.8\% of the surviving errors, with undefined names (B-2) alone rising from 9.0\% to 27.5\%, because a rendering cannot say which API exists.
On the resource side, missing resources (E) grow in absolute terms from 94 to 142 (S1 $\rightarrow$ S2): the invented names are almost exclusively icons; the image makes the model more faithful to what it sees, so it dutifully references the very icons in the screenshot, for which no resource of the same name exists in the target project. A typical case is the settings icon: the source KJC code uses the built-in \texttt{Icons.Filled.Settings} that requires no resource file, and the model renders it in the ArkTS form \texttt{Image(\$r('app.media.ic\_settings'))}; the syntax is right, but the referenced file does not exist, so hvigor rejects the file with \emph{Unknown resource name 'ic\_settings'}.
Fidelity to the screenshot thus creates grounding failures. On SwiftUI, where nonexistent APIs  (B) and wrong usage (C) jointly account for 72.4\% of S1 errors and broken structure (A) takes only 13.5\%, the image brings no relief: the total even rises slightly, from 3,531 to 3,693 (S1 $\rightarrow$ S2), and the duplicate share (A-2) climbs from 1.1\% to 10.8\%.
Hence CSR barely moves, reaching only 2.00\%/0.00\%: structure alone was never the binding constraint in either translation direction.

\begin{boxK}
\small \faIcon{pencil-alt} \textbf{Finding 5:}
The image transmits structure, not knowledge: it collapses structural debris on KJC, cutting total errors by 65\%, yet leaves CSR at 2.00\%/0.00\% on KJC/SwiftUI-to-ArkUI translation, and by making the model more faithful to the icons it sees, it even enlarges the resource-grounding gap that no common strategies can fill.
\end{boxK}

\begin{figure}[t]
    \centering

    \begin{subfigure}[t]{0.49\linewidth}
        \centering
        \includegraphics[height=2.35cm]{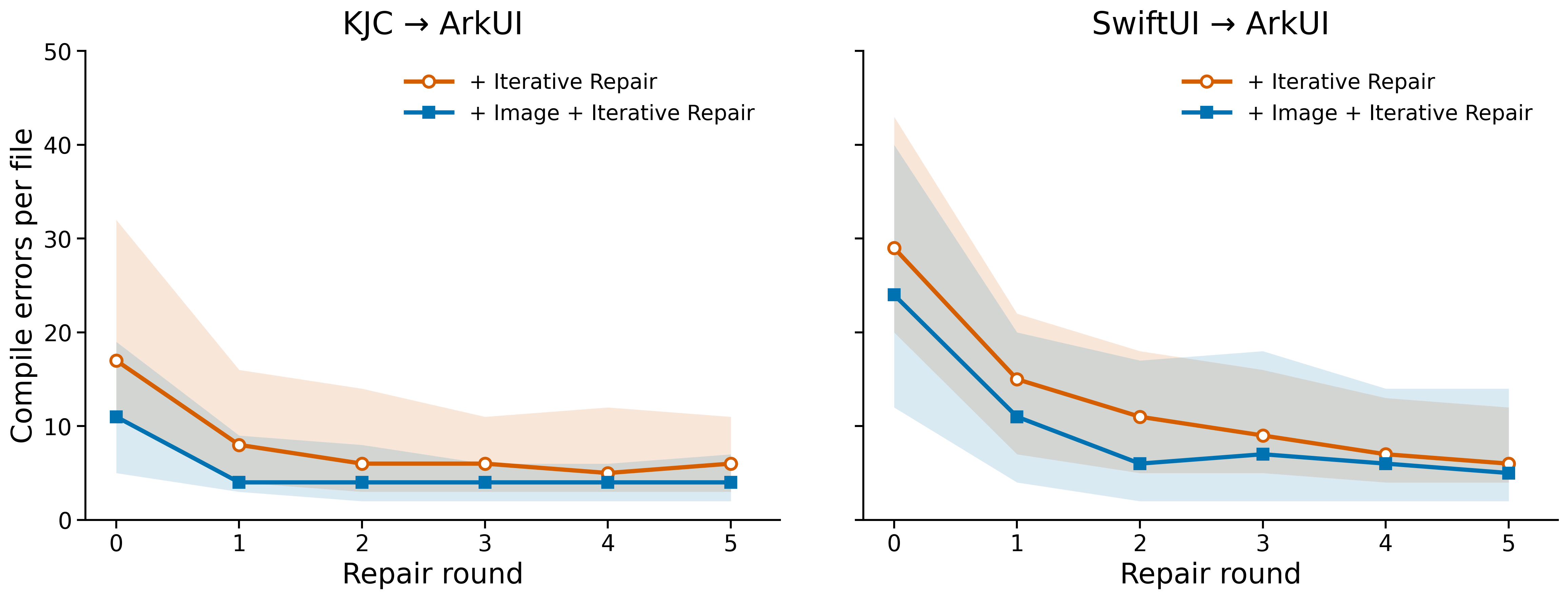}
        \caption{\footnotesize Compile errors within 5 repair rounds}
        \label{fig:rq3_errors}
    \end{subfigure}
    \hfill
    \begin{subfigure}[t]{0.49\linewidth}
        \centering
        \includegraphics[height=2.35cm]{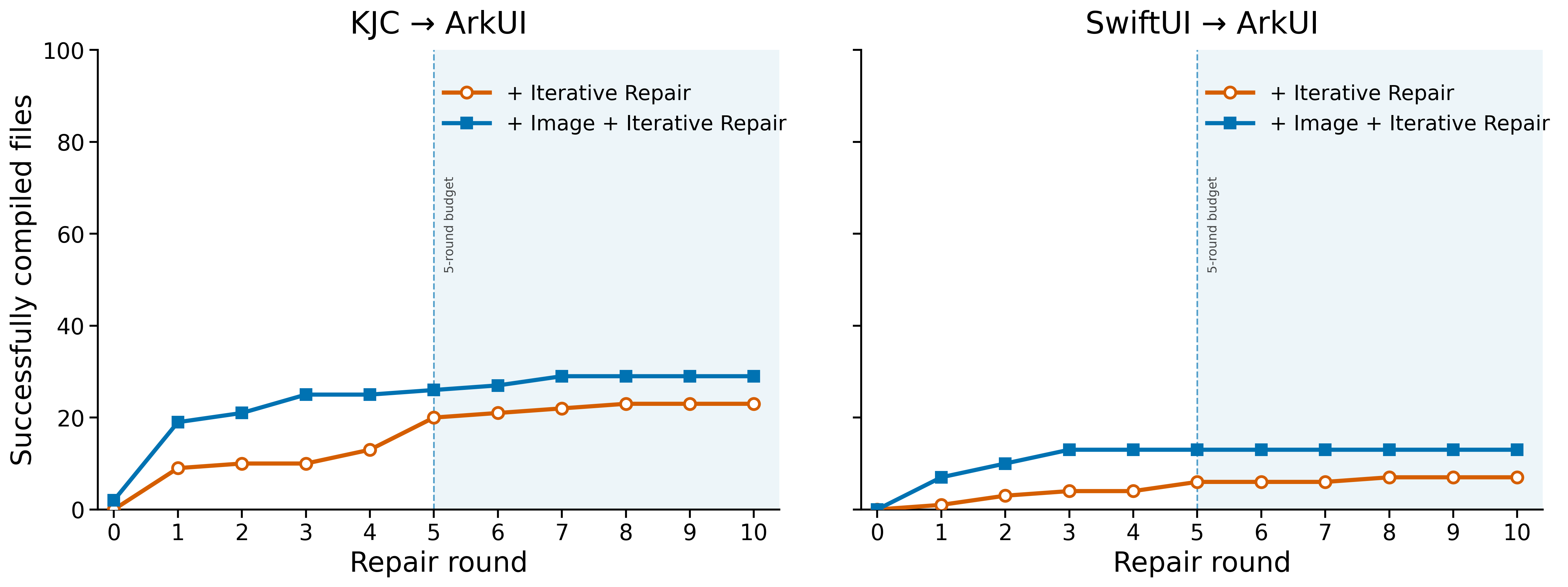}
        \caption{\footnotesize Successful compilations over 10 repair rounds}
        \label{fig:rq3_pass10}
    \end{subfigure}

    \vspace{-0.6em}
    \caption{Effect of repair iterations on compilation outcomes}
    \label{fig:rq3}
    \vspace{-1.2em}
\end{figure}

\textbf{Iteration fixes what needs no knowledge, then degenerates.} The first repair round removes the structural family almost completely: on KJC, duplicate declarations (A-2) fall from 58.9\% in S1 to nearly zero in the S3 final state, and the per-round logs show the mean error count per file falling from 58.3 to 11.6 and the total from 9,361 to 2,655. The family requires no ArkUI knowledge to fix. A duplicate declaration is repaired by deleting the line the compiler points at.
Everything that remains needs knowledge the model lacks, and there the loop has nothing to act on. From round two on, both the level and the spread of errors freeze (Figure~\ref{fig:rq3}(a)). The lines show the median error counts, and the shaded bands indicate the interquartile range (25th--75th percentiles), computed over files that have not yet compiled successfully at each round. The loop's behavior shows why it cannot progress. Carried-over error signatures stop being fixed: the persisted share climbs from 80\% to 91\% while per-round fixes collapse from 370 to 31, so the loop keeps re-feeding exactly the errors it has already failed on. Whole-file rewrites destabilize what repair has settled: 73--85\% of files oscillate, with the error count rising in some round, and duplicates (A-2), eliminated on KJC by round four, reappear in the deeper states. %, reaching 4.3\% on SwiftUI at round five.
Most tellingly, 31\%/23\% of repair rounds resubmit byte-identical code, climbing to 51\%/53\% by round five: iterative repair degenerates into repetition.

The budget itself is not the binding constraint. Figure~\ref{fig:rq3}(b) shows the number of successfully compiled files plateauing after round five: extending the loop from five to ten repair rounds rescues only 3 of the 80 remaining KJC files and 1 of the 94 SwiftUI files under iterative repair alone; under the combined strategy, 3 of 74 KJC and none of 87 SwiftUI files.

% \begin{figure}[t]
%   \centering
%   \includegraphics[width=0.5\linewidth]{fig_rq3.png}
%   \vspace{-1.3em}
%   \caption{Compile errors per file across the iteration budget}
%   \vspace{-1.7em}
%   \label{fig:rq3}
% \end{figure}

\begin{boxK}
\small \faIcon{pencil-alt} \textbf{Finding 6:}
Errors that require no ArkUI knowledge to fix are resolved within the first repair round; the rest stall as a knowledge gap, not a budget limit: 91\% of error signatures persist, half of the repair rounds resubmit byte-identical code, and doubling the budget to ten rounds rescues almost nothing.
\end{boxK}

\textbf{What remains cannot be fixed by common strategies.} The final residue is stable across configurations: nonexistent APIs (B) remain the largest family at 31.6--45.0\%, and missing resources (E) are the only family whose absolute count grows. For these two families, the diagnostic does not contain the remedy. For a duplicate declaration (A-2), by contrast, the compiler output specifies the fix: delete the flagged line. For a nonexistent API, it only reports that the attribute does not exist; identifying the correct attribute requires ArkUI knowledge that the model lacks. For a missing resource, it only reports that the name is unknown.  %the valid names exist only in the target environment, which the model never sees. 
Repair rounds therefore cannot reduce these two families, and that is why none of the strategies exceeds 26.00\%/13.00\% CSR.

\begin{boxK}
\small \faIcon{pencil-alt} \textbf{Finding 7:}
The stalled residue cannot be closed by common strategies: nonexistent APIs (B) remain the largest family in every final state, missing resources (E) are the only family whose absolute count grows, and neither family's diagnostic contains its remedy.
\end{boxK}

\subsection{RQ4: How Far Does \textit{ArkTrans} Lift the Migration Capability of Lightweight LLMs?}
\label{RQ4}
% Throughout this subsection, the \textit{best common strategy} denotes the strongest configuration of RQ2 on the same model, i.e., combining the reference image with compiler iterative repair.
Table~\ref{tab:rq4_ablation} presents the results of the full pipeline and its component ablation. With \textit{ArkTrans}, Qwen3.6-27B now reaches 70.00\% CSR on KJC and 64.00\% on SwiftUI, consuming 13,805 and 13,675 tokens per file, respectively, which is below the 21,758 and 18,231 tokens per file that the best common strategy requires on the same model. The model enters the repair loop at 4.8 average errors on KJC and 3.7 on SwiftUI, rather than 58.3 and 35.3 under direct translation, and converges quickly: before entering the loop, the skeleton-guided generation already compiles 32.00\%/23.00\%, and the first repair round lifts this to 61.00\%/47.00\%.

\begin{boxK}
\small \faIcon{pencil-alt} \textbf{Finding 8:}
\textit{ArkTrans} turns the lightweight model into a usable translator: the full pipeline reaches 70.00\%/64.00\% CSR with substantial visual fidelity at 13,805/13,675 tokens per file, against 0\% in direct translation and 26.00\%/13.00\% for the best common strategy, at a lower token cost.
\end{boxK}

\textbf{Every component is necessary, and each answers one failure
family.} Removing the skeleton is the most damaging ablation: CSR falls from 70.00\% to 47.00\% on KJC and from 64.00\% to 43.00\% on SwiftUI, as the model must again manufacture structure and valid API names on its own, which RQ3 showed feedback cannot reliably supply. This variant still outperforms the best common strategy of RQ2 by 21 points on KJC and 30 on SwiftUI; it runs the same repair loop with the same reference screenshot, so the difference is the one-shot example, which injects the ArkTS usage knowledge that no diagnostic contains.
Removing repair freezes the pipeline at the single generation before the loop, which compiles 32.00\%/23.00\%; iteration therefore contributes 38 points on KJC and 41 on SwiftUI, most of them within the first round.
The reference image plays a small but essential role in compilation, 69.00\% versus 70.00\% on KJC, yet removing it drops Color from 43.17 to 29.72 and Text from 64.58 to 60.13, and on SwiftUI both fall together, CSR to 47.00\% and Color to 20.52. Once the skeleton supplies structure, the image anchors visual properties, exactly the division of labor that RQ3 predicts.

\begin{table*}[htbp]

\centering
\setlength{\abovecaptionskip}{0.1cm}
\caption{Performance of \textit{ArkTrans} and its component ablation}
\label{tab:rq4_ablation}
\renewcommand{\arraystretch}{0.92}%
\small
\resizebox{\linewidth}{!}{
\begin{tabular}{@{}llcccccccc@{}}
\toprule
\multirow{2}{*}{\textbf{Sou. PL}} & \multirow{2}{*}{\textbf{Configuration}} & \multirow{2}{*}{\textbf{CSR (\%)}} & \multirow{2}{*}{\textbf{TK}} & \multicolumn{2}{c}{\textbf{Visual Fidelity (G)}} & \multicolumn{4}{c}{\textbf{Visual Fidelity (L)}} \\ \cmidrule(lr){5-6} \cmidrule(l){7-10}
 &  &  &  & \textbf{CH} & \textbf{CLIP} & \textbf{Pos} & \textbf{Size} & \textbf{Color} & \textbf{Text} \\ \midrule
\multirow{6}{*}{KJC}
 & Direct (RQ1) & 0.00 & 6,132 & 0.00 & 0.00 & 0.00 & 0.00 & 0.00 & 0.00 \\
 & Best common strategy (RQ2: + Image + Repair) & 26.00 & 21,758 & 18.32 & 23.42 & 19.32 & 18.41 & 15.79 & 22.83 \\
 & w/o Skeleton & 47.00 & 19,923 & 31.07 & 44.74 & 36.06 & 37.31 & 31.44 & 45.56 \\
 & w/o Repair & 32.00 & 6,334 & 21.04 & 30.33 & 24.54 & 25.27 & 21.10 & 30.53 \\
 & w/o Image & 69.00 & 13,136 & 43.36 & 58.41 & 39.39 & 40.94 & 29.72 & 60.13 \\
 & \textit{ArkTrans} & \textbf{70.00} & 13,805 & \textbf{45.16} & \textbf{64.51} & \textbf{49.60} & \textbf{51.59} & \textbf{43.17} & \textbf{64.58} \\ \midrule
\multirow{6}{*}{SwiftUI}
 & Direct (RQ1) & 0.00 & 5,992 & 0.00 & 0.00 & 0.00 & 0.00 & 0.00 & 0.00 \\
 & Best common strategy (RQ2: + Image + Repair) & 13.00 & 18,231 & 7.02 & 12.26 & 9.83 & 10.52 & 8.74 & 11.73 \\
 & w/o Skeleton & 43.00 & 25,381 & 28.99 & 38.07 & 29.24 & 29.57 & 25.84 & 37.63 \\
 & w/o Repair & 23.00 & 6,509 & 15.60 & 21.45 & 17.80 & 18.21 & 15.99 & 22.47 \\
 & w/o Image & 47.00 & 16,627 & 29.70 & 40.09 & 27.08 & 27.82 & 20.52 & 42.07 \\
 & \textit{ArkTrans} & \textbf{64.00} & 13,675 & \textbf{41.40} & \textbf{59.70} & \textbf{49.07} & \textbf{48.84} & \textbf{40.47} & \textbf{61.44} \\
\bottomrule
\end{tabular}
}
\begin{tablenotes}
    \footnotesize
    \item \textit{Note:} \textbf{TK} represents the average consumption of tokens per page.
    \vspace{-1.0em}
\end{tablenotes}
\footnotesize
\raggedright
\normalsize

\end{table*}

\begin{boxK} \small \faIcon{pencil-alt} \textbf{Finding 9:} 
Each component is necessary and effective in fostering the migration performance of ArkTrans: without the skeleton, CSR falls to 47.00\%/43.00\%; without repair, to 32.00\%/23.00\%; without the image, KJC loses only color fidelity while SwiftUI loses both compilation and color.

% , with CSR down to 47.00\% and Color to 20.52
\end{boxK}

\textbf{The advantage persists on real-world pages.}
Table~\ref{tab:rq4_realworld} reports the evaluation on 52 UI pages from nine open-source mobile apps, six Android and three iOS, which shift the scale in both directions: the KJC pages average 674.5 lines of code, 2.3 times the benchmark's 291.1, and SwiftUI 145.1, two thirds of its 221.8. The best common strategy collapses to 12.50\%/15.00\% CSR at 25,052/18,231 tokens per file. 
\textit{ArkTrans} attains 65.63\%/80.00\% CSR on the same pages, more than five times the best common strategy, while consuming only 17,423/6,520 tokens per file. Visual fidelity transfers in step, with CH at 9.84/54.21, CLIP at 60.29/74.76, Pos at 45.39/60.58, Size at 47.81/66.09, Color at 33.75/43.80, and Text at 58.53/72.55.

% \begin{table}[htbp]
% \vspace{-1.1em}
% \centering
% \setlength{\abovecaptionskip}{0.1cm}
% \caption{Real-world evaluation from nine open-source apps}
% \label{tab:rq4_realworld}
% \renewcommand{\arraystretch}{0.92}%
% \scriptsize
% \resizebox{\linewidth}{!}{
% \begin{tabular}{@{}llccccccc@{}}
% \toprule
% \multirow{2}{*}{\textbf{Sou. PL}} & \multirow{2}{*}{\textbf{Configuration}} & \multirow{2}{*}{\textbf{CSR (\%)}} & \multicolumn{2}{c}{\textbf{Visual Fidelity (G)}} & \multicolumn{4}{c}{\textbf{Visual Fidelity (L)}} \\ \cmidrule(lr){4-5} \cmidrule(l){6-9}
%  &  &  & \textbf{CH} & \textbf{CLIP} & \textbf{Pos} & \textbf{Size} & \textbf{Color} & \textbf{Text} \\ \midrule
% \multirow{2}{*}{KJC}
%  & Best common strategy (RQ2: + Image + Iterative Repair) & 12.50 & 2.75 & 11.16 & 8.36 & 8.55 & 7.64 & 12.05 \\
%  & \textbf{\textit{ArkTrans}} & \textbf{65.63} & \textbf{9.84} & \textbf{60.29} & \textbf{45.39} & \textbf{47.81} & \textbf{33.75} & \textbf{58.53} \\ \midrule
% \multirow{2}{*}{SwiftUI}
%  & Best common strategy (RQ2: + Image + Iterative Repair) & 15.00 & 3.62 & 14.02 & 5.99 & 8.34 & 6.15 & 9.69 \\
%  & \textbf{\textit{ArkTrans}} & \textbf{80.00} & \textbf{54.21} & \textbf{74.76} & \textbf{60.58} & \textbf{66.09} & \textbf{43.80} & \textbf{72.55} \\
% \bottomrule
% \end{tabular}
% }
% \vspace{-2.2em}
% \end{table}

\begin{boxK}
\small \faIcon{pencil-alt} \textbf{Finding 10:}
On real-world pages,
% disjoint from the benchmark,
\textit{ArkTrans} attains 65.63\%/80.00\% CSR at a lower token cost for KJC/SwiftUI-to-ArkUI translation, while the best common strategy collapses to 12.50\%/15.00\%.
\end{boxK}

\begin{table}[htbp]
\vspace{-1.1em}
\centering
\setlength{\abovecaptionskip}{0.1cm}
\caption{Real-world evaluation from nine open-source apps}
\label{tab:rq4_realworld}
\renewcommand{\arraystretch}{0.92}%
\scriptsize
\resizebox{\linewidth}{!}{
\begin{tabular}{@{}llcccccccc@{}}
\toprule
\multirow{2}{*}{\textbf{Sou. PL}} & \multirow{2}{*}{\textbf{Configuration}} & \multirow{2}{*}{\textbf{CSR (\%)}} & \multirow{2}{*}{\textbf{TK}} & \multicolumn{2}{c}{\textbf{Visual Fidelity (G)}} & \multicolumn{4}{c}{\textbf{Visual Fidelity (L)}} \\ \cmidrule(lr){5-6} \cmidrule(l){7-10}
 &  &  &  & \textbf{CH} & \textbf{CLIP} & \textbf{Pos} & \textbf{Size} & \textbf{Color} & \textbf{Text} \\ \midrule
\multirow{2}{*}{KJC}
 & Best common strategy (RQ2: + Image + Repair) & 12.50 & 25,052 & 2.75 & 11.16 & 8.36 & 8.55 & 7.64 & 12.05 \\
 & \textbf{\textit{ArkTrans}} & \textbf{65.63} & \textbf{17,423} & \textbf{9.84} & \textbf{60.29} & \textbf{45.39} & \textbf{47.81} & \textbf{33.75} & \textbf{58.53} \\ \midrule
\multirow{2}{*}{SwiftUI}
 & Best common strategy (RQ2: + Image + Repair) & 15.00 & 18,231 & 3.62 & 14.02 & 5.99 & 8.34 & 6.15 & 9.69 \\
 & \textbf{\textit{ArkTrans}} & \textbf{80.00} & \textbf{6,520} & \textbf{54.21} & \textbf{74.76} & \textbf{60.58} & \textbf{66.09} & \textbf{43.80} & \textbf{72.55} \\
\bottomrule
\end{tabular}
}
\begin{tablenotes}
    \footnotesize
    \item \textit{Note:} \textbf{TK} represents the average consumption of tokens per page.
    \vspace{-1.0em}
\end{tablenotes}
\footnotesize
\raggedright
\normalsize
\vspace{-0.6em}
\end{table}

\section{Threats to Validity}

\textbf{External Validity} concerns the generality of \textit{ArkTrans}. First, regarding PL generality, our evaluation covers KJC and SwiftUI, the two most widely adopted PLs for Android and iOS, and hence the most prevalent mobile UI scenarios in practice. Second, regarding model dependency, RQ4 evaluates \textit{ArkTrans} on a single lightweight backbone. The choice is conservative: Qwen3.6-27B is among the strongest open-weight multimodal models in its class (Section~\ref{RQ2}), so its failures under direct translation and generic information high-bound those of the entire lightweight class; RQ1--RQ2 further show that the payoff of common strategies is monotone in base capability across the five evaluated models, and every component is model-agnostic: the skeleton is deterministic, and the repair loop interacts with the model only through prompts and compiler diagnostics.
Another threat is the dataset scope: our layouts may not cover all industrial-scale edge cases, but the benchmark diversity shown in Table \ref{tab:code_comparison} and the 52 out-of-benchmark real-world pages in RQ4 probe the robustness of \textit{ArkTrans} beyond curated data.

\noindent\textbf{Internal Validity} relates to potential biases in experimental design. 
% To ensure fair comparison, we established a "Direct LLM" baseline to isolate the impact of ArkUI Skeleton and in-context learning. 
% We maintained identical hyper-parameters (e.g., temperature) across all models to minimize stochastic variance. 
To mitigate implementation errors in deterministic skeleton construction, we manually verified the component mapping dictionary and the generated skeletons against the source UI trees. Regarding data leakage, the risk is minimal as our evaluation benchmark is self-constructed and is not included in the pre-training data of the evaluated LLMs. Although the 52 real-world pages originate from nine open-source apps whose source code is public, no ArkUI ports of these apps exist. Additionally, \cite{yang2026contamination} revealed that functionally equivalent code files of different PLs are normally unpaired in LLMs' pre-training, thus the code translation task is not significantly affected by data leakage. Finally, we have released our complete replication package to promote transparency and reproducibility.

\noindent\textbf{Construct Validity} concerns the evaluation quality. We adopt a multi-dimensional protocol in assessment: (1) \textit{Code executability} is measured via Compilation Success Rate (CSR); (2) \textit{Visual fidelity} is captured through both global metrics (CLIP and CH) and Local Metrics (Position, Size, Color, and Text) of various dimensions. As such, the above holistic evaluation system ensures our measurements align with the practical requirements of functional and aesthetic UI porting.

\section{Related Work}

\subsection{LLMs for Code Translation}
% 分两类，small-context & large-context
Recent work has explored the use of large language models for automated code generation and translation~\cite{roziere2020unsupervised,nashid2023retrieval,chen2021evaluating,zhao2026survey}. Early studies mainly focused on fine-grained translation tasks, where individual functions or methods are translated independently. For example, benchmarks such as MultiPL-E~\cite{cassano2022multipl} provide multilingual datasets for evaluating function-level code generation and translation. Pan et al.~\cite{pan2024lost} further conducted an empirical study of bugs introduced by LLMs during code translation, highlighting the reliability challenges of LLM-based translation systems. Other studies have attempted to improve the performance of function-level code translation through various strategies, such as leveraging test cases for validation (UniTrans~\cite{yang2024exploring}), applying multi-agent frameworks to iteratively repair translations (TransAgent~\cite{10.1145/3797099}), developing specialized translation models (SteloCoder~\cite{pan2023stelocoder}), incorporating external knowledge sources such as API information~\cite{wang2025apirat}, or introducing general correction modules to refine translated code (Rectifier~\cite{yin2024rectifier}). However, function-level translation tools often struggle to capture dependencies across functions and modules, which makes them difficult to apply to real-world software systems. To address this limitation, recent work has explored coarse-grained translation that operates on larger program units, such as classes or entire repositories \cite{xue2025classeval,ibrahimzada2025alphatrans}. More recent approaches further consider repository-level translation and validation across multiple files and modules, including methods that leverage program skeletons to capture the high-level structure of programs during translation~\cite{ibrahimzada2025alphatrans,ke2025advancing,wang2025program}. Inspired by these studies, our work adopts a skeleton-based representation to model the structural organization of UI components during translation.

However, most existing studies focus on backend logic, leaving automated translation across modern declarative UI frameworks underexplored; this work addresses the gap by porting SwiftUI/KJC-based UI frameworks to ArkUI with a dedicated parallel benchmark.

%\subsection{UI Coding Automation}
%Prior work on UI automation mainly studies design-to-code generation~\cite{gui2024vision2ui, wan2024automatically, xiao2024prototype2code, xu2021image2emmet}. Early research translates UI design images into GUI skeletons~\cite{chen2018ui} to bootstrap implementation. More recent work extends this line to declarative UI generation using multimodal models and structured intermediate representations. For example, DeclarUI~\cite{zhou2025declarui} generates declarative UI code from designs, while Design2Code~\cite{si2025design2code} serves as a benchmark for evaluating multimodal front-end code generation from screenshots. Another line of work studies cross-platform UI migration. GUIMigrator~\cite{gao2024rule} migrates Android UI to SwiftUI through skeleton trees, translation rules, and target templates, and UITrans~\cite{gong2025uitrans} translates Android XML layouts into HarmonyOS ArkUI with an LLM-based multi-agent framework. However, these approaches either rely heavily on rule- or template-based transformations, or focus on limited framework pairs with platform-specific translation pipelines.

%To our knowledge, little work studies UI migration as a code translation problem between modern declarative UI frameworks. In this paper, we present the first study on translating SwiftUI and KJC-based UI frameworks to ArkUI.

\subsection{UI Coding Automation}

Prior work on UI automation has predominantly focused on design-to-code generation~\cite{gui2024vision2ui, wan2024automatically, xiao2024prototype2code, xu2021image2emmet}, where UI designs or screenshots are translated into executable code. Early research~\cite{chen2018ui} bootstrapped implementation by translating UI design images into GUI skeletons, while more recent studies~\cite{si2025design2code, zhou2025declarui} extend this to declarative UI generation using multimodal LLMs and structured intermediate representations. To handle complex UIs, another line of work decomposes the task into structural subproblems. For example, DCGen~\cite{wan2024automatically} segments a UI screenshot into components and merges their code snippets into a template, UICopilot~\cite{gui2025uicopilot} derives a DOM skeleton from the design hierarchy and generates code for each subtree, and DesignCoder~\cite{chen2026designcoder} refines generation skeletons through iterative self-correction before merging the generated components. 

However, such decomposition relies on visual heuristics to decide what to split and judges the output mainly with vision-based similarity, while compared to the extensive research on design-to-code, code-to-code UI migration across platforms remains understudied. Existing efforts like GUIMigrator~\cite{gao2024rule} and UITrans~\cite{gong2025uitrans} either do not target ArkUI or still migrate from imperative layouts (e.g., Android XML) rather than modern declarative frameworks. To our knowledge, \textit{ArkTrans} is the first to address UI migration as a direct code translation problem between modern declarative frameworks, specifically porting KJC and SwiftUI to ArkUI.

% Prior work on UI automation has predominantly focused on design-to-code generation~\cite{gui2024vision2ui, wan2024automatically, xiao2024prototype2code, xu2021image2emmet}, where UI designs or screenshots are translated into executable code. Early research~\cite{chen2018ui} often bootstrap implementation by translating UI design images into GUI skeletons, while more recent studies~\cite{si2025design2code, zhou2025declarui} extend this line to declarative UI generation using multimodal models and structured intermediate representations.

% However, compared to the extensive research on design-to-code, code-to-code UI migration across platforms remains understudied. Existing efforts like GUIMigrator~\cite{gao2024rule} and UITrans~\cite{gong2025uitrans} either do not target the HarmonyOS ecosystem or still focus on migrating from outdated imperative layouts (e.g., Android XML) rather than modern declarative frameworks. To our knowledge, \textit{ArkTrans} is the first to address UI migration as a direct code translation problem between modern declarative frameworks, specifically porting KJC and SwiftUI to ArkUI.

\section{Conclusion}

%This work introduces \textit{ArkTrans}, the first skeleton-guided LLM approach for declarative UI migration, and establishes a dedicated parallel benchmark of 100 file-level samples. %Experiments with SOTA LLMs demonstrate that \textit{ArkTrans} significantly outperforms direct and one-shot baselines, achieving a CSR of up to 90.67\% where baselines failed to produce any compilable code. This highlights the necessity of structured skeleton awareness in UI translation. 
%Experiments show that on the lightweight LLM, \textit{ArkTrans} raises the CSR from 0\% to 70.00\% on KJC and 64.00\% on SwiftUI, highlighting the necessity of structured skeleton awareness in UI translation.  Furthermore, our failure-mechanism analysis and component ablation provide practical insights for designing robust UI-aware code translators to enrich the HarmonyOS ecosystem. %Future work will incorporate refined compiler iterative repair and expand the benchmark to include a broader range of UI components.
This work studies KJC/SwiftUI-to-ArkUI UI migration. To achieve this, we build a parallel benchmark of 200 KJC and SwiftUI files across eight scenarios plus 52 real-world mobile pages. Direct translation on diverse LLMs and coding agents yields limited performance, and common strategies significantly lift capable models but fail on lightweight models. Motivated by large-scale manual analysis on root causes,  
% manually analyzing 3,102 samples, deriving per-family regular-expression patterns, and applying them to generalize the resulting failure taxonomy to all 29,593 samples, 
% we find errors rooted in broken structure, nonexistent APIs, and missing resources.  Guided by these roots, 
we propose \textit{ArkTrans} to resolve the limitations of lightweight LLMs.
% which addresses them via Metadata Extraction, Skeleton Construction, LLM-driven Translation, and Compiler-Guided Iterative Repair.  
With Qwen3.6-27B, \textit{ArkTrans} raises CSR to 70.00\% on KJC and 64.00\% on SwiftUI, and attains 65.63\%/80.00\% on real-world pages, outperforming the best common strategy at lower token cost.
Future work will extend \textit{ArkTrans} to repository-level migration and a broader range of UI components.

\section*{Acknowledgments}

This work was partially supported by the National Natural Science Foundation of China (Grant Nos. 62502283, U24B20149), the Natural Science Foundation of Shandong Province (Grant No. ZR2024QF093), and the Young Talent of Lifting Engineering for Science and Technology in Shandong, China (Grant No. SDAST2025QTB031).

\section*{Data Availability}
We open-sourced the data and replication package at \cite{ArkTrans:online}.

% \newpage
\bibliographystyle{ACM-Reference-Format}
\bibliography{software}

@inproceedings{cruzes2011recommended,
  author={Cruzes, Daniela S. and Dyba, Tore},
  booktitle={2011 International Symposium on Empirical Software Engineering and Measurement}, 
  title={Recommended Steps for Thematic Synthesis in Software Engineering}, 
  year={2011},
  volume={},
  number={},
  pages={275-284},
  doi={10.1109/ESEM.2011.36}}

@inproceedings{gui2025uicopilot,
  author = {Gui, Yi and Wan, Yao and Li, Zhen and Zhang, Zhongyi and Chen, Dongping and Zhang, Hongyu and Su, Yi and Chen, Bohua and Zhou, Xing and Jiang, Wenbin and Zhang, Xiangliang},
title = {UICopilot: Automating UI Synthesis via Hierarchical Code Generation from Webpage Designs},
year = {2025},
isbn = {9798400712746},
publisher = {Association for Computing Machinery},
address = {New York, NY, USA},
url = {https://doi.org/10.1145/3696410.3714891},
doi = {10.1145/3696410.3714891},
booktitle = {Proceedings of the ACM on Web Conference 2025},
pages = {1846–1855},
numpages = {10},
location = {Sydney NSW, Australia},
series = {WWW '25}
}

@article{chen2026designcoder,
 author = {Chen, Yunnong and Yu, Xinyu and Ding, Shixian and Zhang, Yingying and Shi, Chengwei and Du, Jingzhou and Chen, Liuqing},
title = {DesignCoder: Hierarchy-aware and self-correcting UI code generation with large language models},
year = {2026},
issue_date = {Oct 2026},
publisher = {Butterworth-Heinemann},
address = {USA},
volume = {198},
number = {C},
issn = {0950-5849},
url = {https://doi.org/10.1016/j.infsof.2026.108214},
doi = {10.1016/j.infsof.2026.108214},
journal = {Inf. Softw. Technol.},
month = aug,
numpages = {16}
}

@inproceedings{qasim2018mobile,
author={Qasim, Iqra and Azam, Farooque and Anwar, Muhammad Waseem and Tufail, Hanny and Qasim, Tehreem},
  booktitle={2018 IEEE 9th Annual Information Technology, Electronics and Mobile Communication Conference (IEMCON)}, 
  title={Mobile User Interface Development Techniques: A Systematic Literature Review}, 
  year={2018},
  volume={},
  number={},
  pages={1029-1034},
  doi={10.1109/IEMCON.2018.8614764}
}

@article{xue2025classeval,
  author = {Xue, Pengyu and Wu, Linhao and Yang, Zhen and Wang, Chengyi and Li, Xiang and Zhang, Yuxiang and Li, Jia and Jin, Ruikai and Pei, Yifei and Shen, Zhaoyan and Lyu, Xiran and Keung, Jacky Wai},
title = {ClassEval-T: Evaluating Large Language Models in Class-Level Code Translation},
year = {2025},
issue_date = {July 2025},
publisher = {Association for Computing Machinery},
address = {New York, NY, USA},
volume = {2},
number = {ISSTA},
url = {https://doi.org/10.1145/3728940},
doi = {10.1145/3728940},
journal = {Proc. ACM Softw. Eng.},
month = jun,
articleno = {ISSTA063},
numpages = {24}
}

@article{yang2026contamination,
  author = {Yang, Zhen and Lin, Hongyi and He, Yifan and Wang, Junqi and Sun, Zeyu and Liu, Shuo and Xu, Jie and Wang, Pengpeng and Yu, Zhongxing and Liang, Qingyuan},
title = {Contamination Means Overestimation? A Fine-Grained Empirical Study in Code Intelligence},
year = {2026},
issue_date = {October 2026},
publisher = {Association for Computing Machinery},
address = {New York, NY, USA},
volume = {3},
number = {ISSTA},
url = {https://doi.org/10.1145/3832279},
doi = {10.1145/3832279},
journal = {Proc. ACM Softw. Eng.},
month = oct,
articleno = {ISSTA188},
numpages = {24}
}

@inproceedings{radford2021learning,
  title = 	 {Learning Transferable Visual Models From Natural Language Supervision},
  author =       {Radford, Alec and Kim, Jong Wook and Hallacy, Chris and Ramesh, Aditya and Goh, Gabriel and Agarwal, Sandhini and Sastry, Girish and Askell, Amanda and Mishkin, Pamela and Clark, Jack and Krueger, Gretchen and Sutskever, Ilya},
  booktitle = 	 {Proceedings of the 38th International Conference on Machine Learning},
  pages = 	 {8748--8763},
  year = 	 {2021},
  editor = 	 {Meila, Marina and Zhang, Tong},
  volume = 	 {139},
  series = 	 {Proceedings of Machine Learning Research},
  month = 	 {18--24 Jul},
  publisher =    {PMLR},
  url = 	 {https://proceedings.mlr.press/v139/radford21a.html}
}

@inproceedings{neubeck2006efficient,
author={Neubeck, A. and Van Gool, L.},
  booktitle={18th International Conference on Pattern Recognition (ICPR'06)}, 
  title={Efficient Non-Maximum Suppression}, 
  year={2006},
  volume={3},
  number={},
  pages={850-855},
  doi={10.1109/ICPR.2006.479}
}

@article{matas2004robust,
  title = {Robust wide-baseline stereo from maximally stable extremal regions},
journal = {Image and Vision Computing},
volume = {22},
number = {10},
pages = {761-767},
year = {2004},
note = {British Machine Vision Computing 2002},
issn = {0262-8856},
doi = {https://doi.org/10.1016/j.imavis.2004.02.006},
url = {https://www.sciencedirect.com/science/article/pii/S0262885604000435},
author = {J Matas and O Chum and M Urban and T Pajdla}
}

@article{canny1986computational,
   author={Canny, John},
  journal={IEEE Transactions on Pattern Analysis and Machine Intelligence}, 
  title={A Computational Approach to Edge Detection}, 
  year={1986},
  volume={PAMI-8},
  number={6},
  pages={679-698},
  doi={10.1109/TPAMI.1986.4767851}
}

@article{haralick1987image,
  author={Haralick, Robert M. and Sternberg, Stanley R. and Zhuang, Xinhua},
  journal={IEEE Transactions on Pattern Analysis and Machine Intelligence}, 
  title={Image Analysis Using Mathematical Morphology}, 
  year={1987},
  volume={PAMI-9},
  number={4},
  pages={532-550},
  doi={10.1109/TPAMI.1987.4767941}
}

@inproceedings{malkoff1997evaluation,
author = {Donald B. Malkoff},
title = {{Evaluation of the Jonker-Volgenant-Castanon (JVC) assignment algorithm for track association}},
volume = {3068},
booktitle = {Signal Processing, Sensor Fusion, and Target Recognition VI},
editor = {Ivan Kadar},
organization = {International Society for Optics and Photonics},
publisher = {SPIE},
pages = {228 -- 239},
year = {1997},
doi = {10.1117/12.280801},
URL = {https://doi.org/10.1117/12.280801}
}

@article{heckbert1982color,
  author = {Heckbert, Paul},
title = {Color image quantization for frame buffer display},
year = {1982},
issue_date = {July 1982},
publisher = {Association for Computing Machinery},
address = {New York, NY, USA},
volume = {16},
number = {3},
issn = {0097-8930},
url = {https://doi.org/10.1145/965145.801294},
doi = {10.1145/965145.801294},
journal = {SIGGRAPH Comput. Graph.},
month = jul,
pages = {297–307},
numpages = {11}
}

@misc{openai,
  author = {{OpenAI}},
  title  = {{GPT-5.5}},
  year   = {2026},
  url    = {https://developers.openai.com/api/docs/models/gpt-5.5}
}

@misc{anthropic,
  author = {{Anthropic}},
  title  = {Introducing {Claude Sonnet 4.6}},
  year   = {2026},
  url    = {https://www.anthropic.com/news/claude-sonnet-4-6}
}

@misc{deepseek,
  author = {{DeepSeek}},
  title  = {{DeepSeek-V4-Pro} GA Release},
  year   = {2026},
  url    = {https://api-docs.deepseek.com/news/news260813/}
}

@misc{qwen,
  author = {{Qwen Team}},
  title  = {{Qwen3.6-27B}: Flagship-Level Coding in a {27B} Dense Model},
  year   = {2026},
  month  = apr,
  url    = {https://qwen.ai/blog?id=qwen3.6-27b}
}

@misc{zhipuai2024glm5,
  author = {{Z.ai}},
  title  = {{GLM-5.2}: Built for Long-Horizon Tasks},
  year   = {2026},
  url    = {https://z.ai/blog/glm-5.2}
}

@misc{codearts,
  author       = {{Huawei Cloud}},
  title        = {{CodeArts IDE}},
  url          = {https://www.huaweicloud.com/product/codeartside.html},
  year         = {2026}
}

@inproceedings{sural2002segmentation,
  author={Sural, S. and Gang Qian and Pramanik, S.},
  booktitle={Proceedings. International Conference on Image Processing}, 
  title={Segmentation and histogram generation using the HSV color space for image retrieval}, 
  year={2002},
  volume={2},
  number={},
  pages={II-II},
  doi={10.1109/ICIP.2002.1040019}
}

@misc{easyocr2020,
  author = {{JaidedAI}},
  title  = {{EasyOCR}},
  year   = {2020},
  url    = {https://github.com/JaidedAI/EasyOCR}
}

@article{yang2024exploring,
  author = {Yang, Zhen and Liu, Fang and Yu, Zhongxing and Keung, Jacky Wai and Li, Jia and Liu, Shuo and Hong, Yifan and Ma, Xiaoxue and Jin, Zhi and Li, Ge},
title = {Exploring and Unleashing the Power of Large Language Models in Automated Code Translation},
year = {2024},
issue_date = {July 2024},
publisher = {Association for Computing Machinery},
address = {New York, NY, USA},
volume = {1},
number = {FSE},
url = {https://doi.org/10.1145/3660778},
doi = {10.1145/3660778},
journal = {Proc. ACM Softw. Eng.},
month = jul,
articleno = {71},
numpages = {24}
}

@article{ibrahimzada2025alphatrans,
author = {Ibrahimzada, Ali Reza and Ke, Kaiyao and Pawagi, Mrigank and Abid, Muhammad Salman and Pan, Rangeet and Sinha, Saurabh and Jabbarvand, Reyhaneh},
title = {AlphaTrans: A Neuro-Symbolic Compositional Approach for Repository-Level Code Translation and Validation},
year = {2025},
issue_date = {July 2025},
publisher = {Association for Computing Machinery},
address = {New York, NY, USA},
volume = {2},
number = {FSE},
url = {https://doi.org/10.1145/3729379},
doi = {10.1145/3729379},
journal = {Proc. ACM Softw. Eng.},
month = jun,
articleno = {FSE109},
numpages = {23}

}

@article{xue2026translibeval,
  author = {Xue, Pengyu and Zheng, Kunwu and Yang, Zhen and Pei, Yifei and Wu, Linhao and Dong, Jiahui and Luo, Xiapu and Xiao, Yan and Liu, Fei and Zhang, Yuxuan and Lyu, Xiran and Li, Xianhang and Zhu, Xuanyu and Wang, Chengyi},
title = {TransLibEval: Demystify Large Language Models’ Capability in Third-Party Library-Targeted Code Translation},
year = {2026},
issue_date = {July 2026},
publisher = {Association for Computing Machinery},
address = {New York, NY, USA},
volume = {3},
number = {FSE},
url = {https://doi.org/10.1145/3797137},
doi = {10.1145/3797137},
journal = {Proc. ACM Softw. Eng.},
month = jun,
articleno = {FSE042},
numpages = {23}
}

@misc{ke2025advancing,
  title={MockingBird: Context-aware Repository-level Code Translation with Isolated Validation}, 
  author={Kaiyao Ke and Ali Reza Ibrahimzada and Rangeet Pan and Saurabh Sinha and Reyhaneh Jabbarvand},
  year={2026},
  eprint={2511.21878},
  archivePrefix={arXiv},
  primaryClass={cs.SE},
  url={https://arxiv.org/abs/2511.21878}, 
}

@article{wang2025program,
  author = {Wang, Bo and Li, Tianyu and Li, Ruishi and Mathur, Umang and Saxena, Prateek},
title = {Program Skeletons for Automated Program Translation},
year = {2025},
issue_date = {June 2025},
publisher = {Association for Computing Machinery},
address = {New York, NY, USA},
volume = {9},
number = {PLDI},
url = {https://doi.org/10.1145/3729287},
doi = {10.1145/3729287},
journal = {Proc. ACM Program. Lang.},
month = jun,
articleno = {184},
numpages = {25}
}

@inproceedings{si2025design2code,
  title = "{D}esign2{C}ode: Benchmarking Multimodal Code Generation for Automated Front-End Engineering",
    author = "Si, Chenglei  and
      Zhang, Yanzhe  and
      Li, Ryan  and
      Yang, Zhengyuan  and
      Liu, Ruibo  and
      Yang, Diyi",
    editor = "Chiruzzo, Luis  and
      Ritter, Alan  and
      Wang, Lu",
    booktitle = "Proceedings of the 2025 Conference of the Nations of the {A}mericas Chapter of the {A}ssociation for {C}omputational {L}inguistics: Human Language Technologies (Volume 1: Long Papers)",
    month = apr,
    year = "2025",
    address = "Albuquerque, New Mexico",
    publisher = "Association for Computational Linguistics",
    url = "https://aclanthology.org/2025.naacl-long.199/",
    doi = "10.18653/v1/2025.naacl-long.199",
    pages = "3956--3974",
    ISBN = "979-8-89176-189-6"
}

@misc{cassano2022multipl,
      title={MultiPL-E: A Scalable and Extensible Approach to Benchmarking Neural Code Generation}, 
      author={Federico Cassano and John Gouwar and Daniel Nguyen and Sydney Nguyen and Luna Phipps-Costin and Donald Pinckney and Ming-Ho Yee and Yangtian Zi and Carolyn Jane Anderson and Molly Q Feldman and Arjun Guha and Michael Greenberg and Abhinav Jangda},
      year={2022},
      eprint={2208.08227},
      archivePrefix={arXiv},
      primaryClass={cs.LG},
      url={https://arxiv.org/abs/2208.08227}, 
}

@inproceedings{pan2024lost,
  author = {Pan, Rangeet and Ibrahimzada, Ali Reza and Krishna, Rahul and Sankar, Divya and Wassi, Lambert Pouguem and Merler, Michele and Sobolev, Boris and Pavuluri, Raju and Sinha, Saurabh and Jabbarvand, Reyhaneh},
title = {Lost in Translation: A Study of Bugs Introduced by Large Language Models while Translating Code},
year = {2024},
isbn = {9798400702174},
publisher = {Association for Computing Machinery},
address = {New York, NY, USA},
url = {https://doi.org/10.1145/3597503.3639226},
doi = {10.1145/3597503.3639226},
booktitle = {Proceedings of the IEEE/ACM 46th International Conference on Software Engineering},
articleno = {82},
numpages = {13},
location = {Lisbon, Portugal},
series = {ICSE '24}
}

@article{10.1145/3797099,
author = {Yuan, Zhiqiang and Chen, Weitong and Wang, Hanlin and Peng, Xin and Chen, Zhenpeng and Lou, Yiling},
title = {TransAgent: Enhancing LLM-Based Code Translation via Fine-Grained Execution Alignment},
year = {2026},
issue_date = {July 2026},
publisher = {Association for Computing Machinery},
address = {New York, NY, USA},
volume = {3},
number = {FSE},
url = {https://doi.org/10.1145/3797099},
doi = {10.1145/3797099},
journal = {Proc. ACM Softw. Eng.},
month = jun,
articleno = {FSE071},
numpages = {23}
}

@misc{pan2023stelocoder,
      title={SteloCoder: a Decoder-Only LLM for Multi-Language to Python Code Translation}, 
      author={Jialing Pan and Adrien Sadé and Jin Kim and Eric Soriano and Guillem Sole and Sylvain Flamant},
      year={2023},
      eprint={2310.15539},
      archivePrefix={arXiv},
      primaryClass={cs.CL},
      url={https://arxiv.org/abs/2310.15539}, 
}

@misc{yin2024rectifier,
      title={Rectifier: Code Translation with Corrector via LLMs}, 
      author={Xin Yin and Chao Ni and Tien N. Nguyen and Shaohua Wang and Xiaohu Yang},
      year={2024},
      eprint={2407.07472},
      archivePrefix={arXiv},
      primaryClass={cs.SE},
      url={https://arxiv.org/abs/2407.07472}, 
}

@inproceedings{wang2025apirat,
  author={Wang, Chaofan and Qiu, Guanjie and Gu, Xiaodong and Shen, Beijun},
  booktitle={2025 IEEE 49th Annual Computers, Software, and Applications Conference (COMPSAC)}, 
  title={ApiRAT: Integrating Multi-source API Knowledge for Enhanced Code Translation with LLMs}, 
  year={2025},
  volume={},
  number={},
  pages={1400-1405},
  doi={10.1109/COMPSAC65507.2025.00176}
}

@article{zhao2026survey,
  author = {Wayne Xin Zhao and
            Kun Zhou and
            Junyi Li and
            Tianyi Tang and
            Zican Dong and
            Yupeng Hou and
            Beichen Zhang and
            Yingqian Min and
            Junjie Zhang and
            Peiyu Liu and
            Xiaolei Wang and
            Yifan Du and
            Chen Yang and
            Yushuo Chen and
            Zhipeng Chen and
            Jinhao Jiang and
            Ruiyang Ren and
            Yifan Li and
            Xinyu Tang and
            Zikang Liu and
            Yiwen Hu and
            Jian-Yun Nie and
            Ji-Rong Wen},
  title   = {A Survey of Large Language Models},
  journal = {Frontiers of Computer Science},
  year    = {2026},
  volume  = {20},
  number  = {12},
  pages   = {2012627},
  doi     = {10.1007/s11704-026-60308-3}
}

@inproceedings{nashid2023retrieval,
  author={Nashid, Noor and Sintaha, Mifta and Mesbah, Ali},
  booktitle={2023 IEEE/ACM 45th International Conference on Software Engineering (ICSE)}, 
  title={Retrieval-Based Prompt Selection for Code-Related Few-Shot Learning}, 
  year={2023},
  volume={},
  number={},
  pages={2450-2462},
  doi={10.1109/ICSE48619.2023.00205}
}

@misc{chen2021evaluating,
  title={Evaluating Large Language Models Trained on Code}, 
      author={Mark Chen and Jerry Tworek and Heewoo Jun and Qiming Yuan and Henrique Ponde de Oliveira Pinto and Jared Kaplan and Harri Edwards and Yuri Burda and Nicholas Joseph and Greg Brockman and Alex Ray and Raul Puri and Gretchen Krueger and Michael Petrov and Heidy Khlaaf and Girish Sastry and Pamela Mishkin and Brooke Chan and Scott Gray and Nick Ryder and Mikhail Pavlov and Alethea Power and Lukasz Kaiser and Mohammad Bavarian and Clemens Winter and Philippe Tillet and Felipe Petroski Such and Dave Cummings and Matthias Plappert and Fotios Chantzis and Elizabeth Barnes and Ariel Herbert-Voss and William Hebgen Guss and Alex Nichol and Alex Paino and Nikolas Tezak and Jie Tang and Igor Babuschkin and Suchir Balaji and Shantanu Jain and William Saunders and Christopher Hesse and Andrew N. Carr and Jan Leike and Josh Achiam and Vedant Misra and Evan Morikawa and Alec Radford and Matthew Knight and Miles Brundage and Mira Murati and Katie Mayer and Peter Welinder and Bob McGrew and Dario Amodei and Sam McCandlish and Ilya Sutskever and Wojciech Zaremba},
      year={2021},
      eprint={2107.03374},
      archivePrefix={arXiv},
      primaryClass={cs.LG},
      url={https://arxiv.org/abs/2107.03374}, 
}

@inproceedings{roziere2020unsupervised,
  author = {Roziere, Baptiste and Lachaux, Marie-Anne and Chanussot, Lowik and Lample, Guillaume},
 booktitle = {Advances in Neural Information Processing Systems},
 editor = {H. Larochelle and M. Ranzato and R. Hadsell and M.F. Balcan and H. Lin},
 pages = {20601--20611},
 publisher = {Curran Associates, Inc.},
 title = {Unsupervised Translation of Programming Languages},
 url = {https://proceedings.neurips.cc/paper_files/paper/2020/file/ed23fbf18c2cd35f8c7f8de44f85c08d-Paper.pdf},
 volume = {33},
 year = {2020}
}

@book{sadun2013ios,
  title={iOS Drawing: Practical UIKit Solutions},
  author={Sadun, Erica},
  year={2013},
  publisher={Addison-Wesley}
}

@incollection{jackson2014introduction,
  author="Jackson, Wallace",
title="Introduction to XML: Defining an Android App, Its Design, and Constants",
bookTitle="Android Apps for Absolute Beginners",
year="2014",
publisher="Apress",
address="Berkeley, CA",
pages="101--130",
isbn="978-1-4842-0019-3",
doi="10.1007/978-1-4842-0019-3_4",
url="https://doi.org/10.1007/978-1-4842-0019-3_4"
}

@inproceedings{gong2025uitrans,
  author = {Gong, Lina and Wang, Chen and Cui, Di and Huang, Yujun and Wei, Mingqiang},
title = {UITrans: Seamless UI Translation from Android to HarmonyOS},
year = {2025},
isbn = {9798400719264},
publisher = {Association for Computing Machinery},
address = {New York, NY, USA},
url = {https://doi.org/10.1145/3755881.3755933},
doi = {10.1145/3755881.3755933},
booktitle = {Proceedings of the 16th International Conference on Internetware},
pages = {142–146},
numpages = {5},
location = {Trondheim, Norway},
series = {Internetware '25}
}

@article{zhou2025porting,
  author = {Zhou, Bo and Shi, Jiaqi and Wang, Ying and Li, Li and Li, Tsz On and Yu, Hai and Zhu, Zhiliang},
title = {Porting Software Libraries to OpenHarmony: Transitioning from TypeScript or JavaScript to ArkTS},
year = {2025},
issue_date = {July 2025},
publisher = {Association for Computing Machinery},
address = {New York, NY, USA},
volume = {2},
number = {ISSTA},
url = {https://doi.org/10.1145/3728941},
doi = {10.1145/3728941},
journal = {Proc. ACM Softw. Eng.},
month = jun,
articleno = {ISSTA064},
numpages = {22}
}

@book{mathias2016swift,
  author    = {Matthew Mathias and John Gallagher},
  title     = {{Swift} Programming: The Big Nerd Ranch Guide},
  publisher = {Big Nerd Ranch Guides},
  year      = {2016},
  edition   = {2},
  isbn      = {9780134610696}
}

@book{jemerov2017kotlin,
  author    = {Dmitry Jemerov and Svetlana Isakova},
  title     = {{Kotlin} in Action},
  publisher = {Manning Publications},
  year      = {2017},
  isbn      = {9781617293290}
}

@book{10522550,
  author    = {Fadda, Michele},
  title     = {An {iOS} Developer's Guide to {SwiftUI}: Design and Build Beautiful Apps Quickly and Easily with Minimum Code},
  publisher = {Packt Publishing},
  year      = {2024},
  isbn      = {9781801813624}
}

@inproceedings{marchenko2023jetpack,
  author    = {S. Marchenko},
  title     = {{Jetpack Compose}: New Approaches to {Android UI} Development},
  booktitle = {Information Technologies and Management in Higher Education and Sciences. Part 2},
  publisher = {Baltija Publishing},
  year      = {2022},
  pages     = {281--284},
  doi       = {10.30525/978-9934-26-277-7-160}
}

@article{zhou2025declarui,
author = {Zhou, Ting and Zhao, Yanjie and Hou, Xinyi and Sun, Xiaoyu and Chen, Kai and Wang, Haoyu},
title = {DeclarUI: Bridging Design and Development with Automated Declarative UI Code Generation},
year = {2025},
issue_date = {July 2025},
publisher = {Association for Computing Machinery},
address = {New York, NY, USA},
volume = {2},
number = {FSE},
url = {https://doi.org/10.1145/3715726},
doi = {10.1145/3715726},
journal = {Proc. ACM Softw. Eng.},
month = jun,
articleno = {FSE011},
numpages = {23}
}

@misc{gao2024rule,
  title={GUIMigrator: Semantics-Preserving Transpilation from Android XML to Compose and SwiftUI}, 
  author={Yi Gao and Xing Hu and Xin Xia and Xiaohu Yang},
  year={2026},
  eprint={2409.16656},
  archivePrefix={arXiv},
  primaryClass={cs.SE},
  url={https://arxiv.org/abs/2409.16656}, 
}

@inproceedings{chen2018ui,
author = {Chen, Chunyang and Su, Ting and Meng, Guozhu and Xing, Zhenchang and Liu, Yang},
title = {From UI design image to GUI skeleton: a neural machine translator to bootstrap mobile GUI implementation},
year = {2018},
isbn = {9781450356381},
publisher = {Association for Computing Machinery},
address = {New York, NY, USA},
url = {https://doi.org/10.1145/3180155.3180240},
doi = {10.1145/3180155.3180240},
booktitle = {Proceedings of the 40th International Conference on Software Engineering},
pages = {665–676},
numpages = {12},
location = {Gothenburg, Sweden},
series = {ICSE '18}
}

@misc{gui2024vision2ui,
    title={WebCode2M: A Real-World Dataset for Code Generation from Webpage Designs}, 
      author={Yi Gui and Zhen Li and Yao Wan and Yemin Shi and Hongyu Zhang and Yi Su and Bohua Chen and Dongping Chen and Siyuan Wu and Xing Zhou and Wenbin Jiang and Hai Jin and Xiangliang Zhang},
      year={2025},
      eprint={2404.06369},
      archivePrefix={arXiv},
      primaryClass={cs.CV},
      doi={https://doi.org/10.1145/3696410.3714889},
      url={https://arxiv.org/abs/2404.06369}, 
}

@article{wan2024automatically,
  title={Divide-and-Conquer: Generating UI Code from Screenshots},
   volume={2},
   ISSN={2994-970X},
   url={http://dx.doi.org/10.1145/3729364},
   DOI={10.1145/3729364},
   number={FSE},
   journal={Proceedings of the ACM on Software Engineering},
   publisher={Association for Computing Machinery (ACM)},
   author={Wan, Yuxuan and Wang, Chaozheng and Dong, Yi and Wang, Wenxuan and Li, Shuqing and Huo, Yintong and Lyu, Michael},
   year={2025},
   month=June, pages={2099–2122} }

@misc{xiao2024prototype2code,
  title={Prototype2Code: End-to-end Front-end Code Generation from UI Design Prototypes}, 
  author={Shuhong Xiao and Yunnong Chen and Jiazhi Li and Liuqing Chen and Lingyun Sun and Tingting Zhou},
  year={2024},
  eprint={2405.04975},
  archivePrefix={arXiv},
  primaryClass={cs.SE},
  url={https://arxiv.org/abs/2405.04975}, 
}

@article{xu2021image2emmet,
author = {Xu, Yong and Bo, Lili and Sun, Xiaobing and Li, Bin and Jiang, Jing and Zhou, Wei},
title = {image2emmet: Automatic code generation from web user interface image},
year = {2021},
issue_date = {August 2021},
publisher = {John Wiley \& Sons, Inc.},
address = {USA},
volume = {33},
number = {8},
issn = {2047-7473},
url = {https://doi.org/10.1002/smr.2369},
doi = {10.1002/smr.2369},
journal = {J. Softw. Evol. Process},
month = aug,
numpages = {21}
}

@misc{ArkTrans:online,
author       = {Anonymous},
  title        = {Data for ArkTrans},
  year         = {2026},
  publisher    = {github},
  url          = {https://github.com/AnonymousAuthor0856/ArkTrans},
  note = "[Online; accessed 2026-11-04]"
}
\end{document}